\documentclass[prb,twocolumn,showpacs,preprintnumbers,amsmath,amssymb,superscriptaddress]{revtex4}
\usepackage{graphicx}
\usepackage{dcolumn}
\usepackage{bm}
\usepackage{psfrag}

\begin{document}
\title{Time-dependent Gutzwiller theory of pairing fluctuations in the Hubbard
model} 
\author{G. Seibold}
\affiliation{Institut f\"ur Physik, BTU Cottbus, PBox 101344,
03013 Cottbus, Germany}
\author{F. Becca}
\affiliation{CNR-INFM Democritos, National Simulation Centre, and SISSA I-34014
Trieste, Italy.}
\author{J. Lorenzana}
\affiliation{Center for Statistical
Mechanics and Complexity, INFM, Dipartimento di Fisica,
Universit\`a di Roma La Sapienza, P. Aldo Moro 2, 00185 Roma, Italy}
\date{\today}
\begin{abstract}
We present a method to compute pairing fluctuations on top of the Gutzwiller 
approximation (GA). Our investigations are based on a charge-rotational 
invariant GA energy functional which is expanded up to second order in the 
pair fluctuations. Equations of motion for the fluctuations lead to a 
renormalized ladder type approximation. Both spectral functions and corrections
to static quantities, like the ground-state energy, are computed.  
The quality of the method is examined for the single-band Hubbard model 
where we compare the dynamical pairing correlations for s- and d-wave 
symmetries with exact diagonalizations and find a significant improvement with
respect to analogous calculations done within the standard Hartree-Fock 
ladder approximation. The technique has potential applications in the theory 
of Auger spectroscopy, superconductivity, and cold atom physics.     
\end{abstract}

\maketitle

\section{Introduction}
The present interest in the physics of strongly correlated fermion systems 
is accompanied by a `revival' of the Gutzwiller wave function 
(GWF)~\cite{GUTZ1} in order to study Hubbard-type models.
In these Hamiltonians, doubly occupied sites contribute with an energy
$U$ to the total energy, though their weight is partially reduced by the
Gutzwiller projector. Originally, the projector was applied to a Slater 
determinant describing a spatially uniform Fermi liquid. Along the
years, states with long-range order have been considered, including 
projected BCS wave functions,~\cite{bask87,kot88,para01,sorella02}
used in the context of high-$T_c$.
 
Analytic evaluations of expectation values using the GWF are only possible
in one~\cite{METZNER1} and infinite dimensions.~\cite{METZNER2,GEBHARD} 
In the latter limit, one recovers the so-called Gutzwiller approximation 
(GA) first introduced by Gutzwiller himself in Ref.~\onlinecite{GUTZ2}.
The GA, later on rederived as a saddle-point within a slave-boson
formulation of the Hubbard model,~\cite{KR} yields an energy functional
for quasiparticles with renormalized hopping amplitude. Therefore, it
offers a simple and intuitive picture for the interplay between
local correlation and electron kinetics and it is used in a variety of fields 
including the description of inhomogeneous states in cuprates,~\cite{lor02} 
band structure calculations,~\cite{buenemann} and the theory of 
He$^3$.~\cite{VOLLHARDT}
   
In the past few years, two of us have developed a scheme which allows
the computation of Gaussian fluctuations on top of the 
GA.~\cite{GOETZ1,GOETZ2,goetz03} Within this so-called time-dependent
GA (TDGA), it is possible to evaluate dynamical correlation functions in the
charge and spin channel, which over a wide parameter regime are
in good agreement with exact diagonalization results and constitute
a significant improvement over the traditional Hartree-Fock plus Random-phase
approximation (HF+RPA).
The TDGA has been used in order to compute dynamical properties of 
inhomogeneous states in cuprates. Results obtained in this way for the optical 
conductivity~\cite{lor03} and magnetic susceptibility~\cite{goe05,goe06} of 
stripe ordered phases have turned out to be in excellent agreement with 
experimental data. 

In the present paper, we generalize the TDGA towards the inclusion
of pairing fluctuations. The significance of such correlations is probably
most prominent in the context of superconductivity where the appearance
of a singularity in the pair susceptibility signals the onset of
Cooper-pair condensation. Here, we aim to study the spectrum of pairing 
correlations in the normal state of the Hubbard model, an issue
which has also been addressed, among others, within exact 
diagonalization~\cite{dag90,becca00} and various Monte Carlo 
methods.~\cite{white89,moreo92,bulut93,zhang97,plekha05,arita06,yana07}
In the particle-hole channel, the coupling to an electric or magnetic field 
yields the optical conductivity and magnetic susceptibility, respectively.
In the particle-particle channel, direct measurements of pairing correlations 
are not so common since they do not couple to a classical field. However, 
principles for the measurement of the pair susceptibility in metals have been 
discussed in Refs.~\onlinecite{ferrell69,scala70,taka71,shenoy74}.
The basic idea is to probe the fluctuating pair field of a metal
in the normal state which is coupled to a superconductor via the
tunnel current-voltage characteristics.
In addition, the pairing correlation function for local pairs is the
main ingredient in the theory of Auger 
spectroscopy~\cite{cini76,sawa77,cin86,ver01} and
also has relevance in the field of ultra-cold atom physics.~\cite{winkler06}

In materials which can be described within Hubbard-type models, a strong 
local repulsion induces so-called antibound states (also known as two-particle 
resonances), above the two-particle continuum. The resonances appear as 
atomic-like features in the Auger spectrum.
In Ref.~\onlinecite{sei08} we have shown that, despite 
its computational simplicity, the TDGA yields an excellent description of the 
two-particle response. In particular, it describes well the 
energy gap between band and antibound states and the relative spectral weight
even far from the dilute limit in contrast to the `bare ladder approximation' 
(BLA), which is restricted to the low-density
regime.~\cite{gal58,kan63} Thus, the TDGA allows one to extend ladder-type
theories much beyond their supposed limit of validity. The result is
an effective ladder theory where quasiparticles get heavier, as usual
due to correlations, and at the same time the effective
interactions between quasiparticles become strongly
renormalized. These vertex and self-energy corrections are consistent
with each other and do not suffer from the pitfalls frequently found in
diagrammatic computations, where an improvement at the level of the
self-energy alone leads to a degradation of the overall performance of
the theory due to the lack of appropriate 
vertex corrections.~\cite{cin86,ver01}

In this paper, besides a thorough derivation of the `pairing TDGA', we extend 
the approach to intersite correlations with extended s-wave and d-wave 
symmetry. We will show that the TDGA yields an effective interaction between 
quasiparticles which is renormalized with respect to the bare $U$ due to
correlations but it does not include enough fluctuation effects to induce
a superconducting instability. This is because we start from a
paramagnetic state treated in the GA which does not have enough
variational freedom to describe the scale $J$ of magnetic
fluctuations. In addition, within the RPA treatment for a paramagnet  
the particle-particle and particle-hole channel are decoupled and do not 
influence each other. For spin-density wave
ground states, the TDGA induces attractive interactions between 
nearest-neighbor pairs which are not present in the HF approach based
on BLA. However, these are not enough to produce a superconducting
instability. Superconductivity due to an electronic mechanism requires
the feedback of particle-hole fluctuations on the particle-particle
channel and this goes beyond our approach. On the other hand, at the
level of spectroscopic quantities this is a minor effect and our approach 
turns out to be in excellent agreement with exact diagonalizations.
Here and below we use the terms ``ladder-type fluctuations'',
``particle-particle RPA fluctuations'' and ``pairing fluctuations''
as synonymous.  

This paper is organized as follows: The formalism is presented in detail
in Sec.~\ref{section:II} where as a first step we derive the 
charge-rotational invariant GA functional from the Hubbard model in 
Sec.~\ref{sec2a}. Based on this functional, we show in Sec.~\ref{sec2b} 
how ladder-type fluctuations can be incorporated into the approach and how 
dynamical pair correlations can be computed. 
Results are presented in Sec.~\ref{section:III} where we first exemplify the
method by means of a two-site model. Then the dynamical pairing correlations
for different symmetries are computed on small clusters and compared with
exact diagonalization results and HF+BLA computations.
Finally, we conclude our investigations in Sec.~\ref{sec:conc}.

\section{Formalism}\label{section:II}

\subsection{Charge-rotationally invariant GA}\label{sec2a}

The starting point is the one-band Hubbard model:
\begin{equation}\label{HM}
H= \sum_{i,j,\sigma} (t_{ij}-\mu \delta_{ij}) c_{i\sigma}^{\dagger}c_{j\sigma} + U\sum_{i}
\hat n_{i\uparrow}\hat n_{i\downarrow},
\end{equation}
where $c_{i\sigma}$ ($c^\dagger_{i\sigma}$) destroys (creates) an electron 
with spin $\sigma$ at site
$i$, and $\hat n_{i\sigma}=c_{i\sigma}^{\dagger}c_{i\sigma}$. $U$ is the
on-site Hubbard repulsion, $t_{ij}$ denotes the hopping parameter between
sites $i$ and $j$ and $\mu$ is the chemical potential.

The Gutzwiller variational wave function can be written as
\begin{equation}\label{gwf}
|\Phi_G\rangle \equiv \prod_i\hat{P}_i|\phi\rangle,
\end{equation}
where $\hat{P}_i$ partially projects out a doubly occupied state at site $i$
from the uncorrelated wave function $|\phi\rangle$.
In the traditional Gutzwiller approach,~\cite{GUTZ1} the latter is a Slater 
determinant and the associated density matrix only contains the normal part:
$$\rho_{ij}^{\sigma\sigma'} \equiv \langle \phi|c_{j\sigma'}^\dagger c_{i\sigma}|\phi\rangle.$$
Here, we will consider a more general formulation in which $|\phi\rangle $ is 
a Bogoliubov vacuum and define the anomalous part of the density matrix 
$$\kappa_{ij} \equiv \langle \phi| c_{j\downarrow}c_{i\uparrow}|
\phi\rangle.$$ 
In the general case the normal part can describe charge-density wave and 
spin-density wave broken symmetries. We will allow the ground state 
to have these broken symmetries but we will assume it is normal so our
saddle point anomalous density matrix  will vanish. We denote quantities 
at the saddle point by a 0, thus $\kappa_{ij}^0=0$. 
The anomalous part is important in order to obtain 
the fluctuations. Indeed, in the following, we consider an external 
time-dependent perturbation which induces pairing fluctuations on 
$|\phi\rangle$, then the instantaneous $|\phi\rangle$ will take a BCS-like 
form.

The charge-rotationally invariant Gutzwiller functional 
for general charge and spin textures is derived in the
Appendix~\ref{AP} exploiting the well known equivalence 
between the slave-boson method and the Gutzwiller approach 
and following previous works.~\cite{FW,bulka1,bulka2,MICNAS,sofo}
Alternatively, one can use a pure Gutzwiller formulation with
an appropriate projector $\hat{P}_i$.~\cite{buene05}
The generalized energy functional for the Hubbard model reads: 
\begin{eqnarray}
F({\rho},{\kappa},{D}) &=& 
\sum_{i,j}
t_{ij} \langle \phi| {\bf \Psi}_i^\dagger {\bf A}_{i} \boldsymbol{\tau}_z
{\bf A}_{j}{\bf \Psi}_j|\phi\rangle \nonumber \\
&-& \mu  \sum_{i}\rho_{ii} + U\sum_{i} D_i, 
\label{EGA}
\end{eqnarray}
which depends on the normal and anomalous parts of the density matrix and
the variational double occupancy parameters $D_i$.  
In Eq.~\eqref{EGA}, $\rho_{ii}\equiv\sum_\sigma \rho_{ii}^{\sigma\sigma}$ 
and we have introduced Nambu notation 
\begin{equation}
{\bf \Psi}_i^\dagger = (c_{i\uparrow}^\dagger , c_{i\downarrow})
\,\,\,\,\,\,  {\bf \Psi}_i = \left(\begin{array}{c} c_{i\uparrow} \\
c_{i\downarrow}^\dagger \end{array}\right).
\end{equation}
It is also convenient to define the following pseudospin vector
\begin{eqnarray}
 J_{i}^x &=& \frac{1}{2}{\bf \Psi}_i^\dagger \boldsymbol{\tau}_x{\bf \Psi}_i=  \frac{1}{2}\left(c_{i\uparrow}^\dagger c_{i\downarrow}^\dagger
+c_{i\downarrow} c_{i\uparrow}\right), \\
J_{i}^y &=& \frac{1}{2}{\bf \Psi}_i^\dagger \boldsymbol{\tau}_y{\bf \Psi}_i=-\frac{i}{2}\left(c_{i\uparrow}^\dagger c_{i\downarrow}^\dagger
-c_{i\downarrow} c_{i\uparrow}\right), \\
J_{i}^z &=& \frac{1}{2}{\bf \Psi}_i^\dagger \boldsymbol{\tau}_z{\bf \Psi}_i=\frac{1}{2}\left(c_{i\uparrow}^\dagger c_{i\uparrow}
+c_{i\downarrow}^\dagger c_{i\downarrow} -1\right), \label{eqjz}
\end{eqnarray}
where $\boldsymbol{\tau}_i$ denote the Pauli matrices. The components of 
${\vec J}_i$ obey the standard commutation relations of a spin $1/2$ algebra.
We use boldface latter to indicate two-component vectors and $2\times2$
matrices in Nambu space, whereas we denote Cartesian vectors by using an arrow. 

The raising and lowering operators are defined as 
\begin{eqnarray}
J_i^+ &=& J_i^x +i J_i^y =  c_{i\uparrow}^\dagger c_{i\downarrow}^\dagger, 
\label{eqjp}\\
J_i^- &=& J_i^x -i J_i^y =  c_{i\downarrow} c_{i\uparrow} \label{eqjm}.
\end{eqnarray}

Within these definitions the matrix ${\bf A}_{i}$ in Eq.~\eqref{EGA} reads as
\begin{equation}\label{amat}
{\bf A}_i= \left( \begin{array}{cc} 
\frac{z_{i\uparrow}+z_{i\downarrow}}{2}
+ \frac{z_{i\uparrow}-z_{i\downarrow}}{2}\frac{\langle J_i^z\rangle}
{\langle J_i\rangle} & 
\frac{\langle J_i^-\rangle }{2 \langle J_i \rangle } [z_{i\uparrow}-z_{i\downarrow}] \\ 
\frac{\langle J_i^+  \rangle}{2\langle J_i \rangle }
[z_{i\uparrow}-z_{i\downarrow}]& 
\frac{z_{i\uparrow}+z_{i\downarrow}}{2}
- \frac{z_{i\uparrow}-z_{i\downarrow}}{2}\frac{\langle J_i^z\rangle}{\langle 
J_i\rangle}
\end{array} \right),
\end{equation}
and all expectation values $\langle \dots \rangle$ refer to the state 
$| \phi\rangle$.
The Gutzwiller renormalization factors are given by
\begin{widetext}
\begin{equation}\label{qfak}
z_{i\sigma}=\frac{\sqrt{D_i-\langle J_i^z\rangle -\langle J_i\rangle}
\sqrt{1/2+\langle J_i^z\rangle-D_i+\sigma \langle S_i^z\rangle}
+\sqrt{1/2+\langle J_i^z\rangle-D_i-\sigma \langle S_i^z\rangle}
\sqrt{D_i-\langle J_i^z\rangle +\langle J_i\rangle}}
{\sqrt{1/4-(\langle J_i\rangle+\langle S_i^z\rangle)^2}}.
\end{equation}
\end{widetext}
Note that in the limit $\langle J_i^\pm\rangle=0$, where the matrix 
${\bf A}_i$ is diagonal, one recovers the standard Gutzwiller energy functional
as derived by Gebhard~\cite{GEBHARD} or Kotliar-Ruckenstein:~\cite{KR}
\begin{equation}
F(\rho,D)= \sum_{i,j} t_{ij\sigma}
z_{i\sigma}z_{j\sigma}\rho_{ji}^{\sigma\sigma}-\mu\sum_i \rho_{ii}
+ U\sum_{i} D_i,
\end{equation}
with the hopping renormalization factors
\begin{equation}\label{qfac}
z_{i\sigma}=\frac{\sqrt{(\rho_{ii}^{\sigma\sigma}-D_i)(1-\rho_{ii}+D_i)}+
\sqrt{(\rho_{ii}^{\bar\sigma\bar\sigma}-D_i)D}_i}{\sqrt{\rho_{ii}^{\sigma\sigma}(1-\rho_{ii}^{\sigma\sigma})}}.
\end{equation}
The minimization of the energy functional of Eq.~\eqref{EGA} leads to the
stationary Gutzwiller wave function and to the associated stationary
uncorrelated state $|\phi_0\rangle$. 

\subsection{Calculation of pair fluctuations around general GA saddle points} 
\label{sec2b}

The energy functional of Eq.~\eqref{EGA} is a convenient starting point
for the calculation of pair excitations on top of unrestricted Gutzwiller 
wave functions. Following the general approach of 
Refs.~\onlinecite{GOETZ1,GOETZ2,goetz03,sei08}, we study the response of the 
system to an external time-dependent perturbation which induces small-amplitude
oscillations in the particle-particle channel:
\begin{equation}
\label{eq:fdt}
{\cal F}(t)=\sum_{i} (f_{ij} e^{-\imath \omega t} c_{i\downarrow} c_{j\uparrow}
+H.c.).
\end{equation}
Correspondingly, we have to expand the energy functional of Eq.~\eqref{EGA}
around the stationary solution up to second order in the density and 
double-occupancy deviations.
As already mentioned, we shall restrict to saddle-point solutions in
the normal state: 
\begin{equation}
\kappa_{ij}^0=\langle J_i^+\rangle_0 = \langle J_i^-\rangle_0 = 0,
\end{equation}
and we remind the reader that a subscript or superscript $0$ indicates
quantities evaluated in the stationary state $|\phi_0\rangle$. Fluctuations 
are defined as $\delta \rho(t) =\rho(t)-\rho_0$, etc.

In the present case, particle-hole (ph) and particle-particle (pp) sectors in 
the expansion are decoupled and one obtains  
\begin{equation} \label{E2}
F=F_0+\mbox{tr}\{h^0 \delta{\rho}\} + \delta F^{\rm ph} 
+ \delta F^{\rm pp}, 
\end{equation}
where we have introduced the Gutzwiller Hamiltonian:~\cite{RING,BLAIZOT}
\begin{equation}
  \label{eq:hgw}
  h_{ij}^{\sigma\sigma'}[\rho,D]=\frac{\partial F} 
{\partial \rho_{ji}^{\sigma'\sigma}}\delta_{\sigma,\sigma'}.
\end{equation}
This coincides with the Kotliar-Ruckenstein Hamiltonian matrix.~\cite{KR} 
In particular, the diagonal elements (in the basis of atomic orbitals) 
coincide with the Lagrange multipliers of the  Kotliar-Ruckenstein 
method, $\Sigma_{i\sigma}$, after adding the chemical potential:
\begin{equation}\label{eq:lagrange}
\Sigma_{i\sigma}=\frac{\partial F}{\partial \rho_{ii}^{\sigma\sigma}}+\mu .
\end{equation}
In Ref.~\onlinecite{sei08} we have interpreted $\Sigma_{i\sigma}$ as a
local GA self-energy.

Since we have included $\mu$ in Eq.~\eqref{HM} the eigenvalues of  
Eq.~\eqref{eq:hgw}, $\xi$, describe the single-particle
excitations with respect to the chemical potential at the GA
level. We denote 
the particle (hole) energies above (below) the Fermi energy by $\xi_{p\sigma}$ 
($\xi_{h\sigma}$) with $\xi_{p\sigma}>0>\xi_{h\sigma}$.

The transformation from real space fermions $c_{i\sigma}$ to 
GA operators is written as
\begin{equation}\label{eq:trafo}
c_{i\sigma} = \sum_p\phi_{i\sigma}(p)a_{p\sigma} + \sum_h\phi_{i\sigma}(h)a_{h\sigma} 
\end{equation}
and the amplitudes $\phi_{i\sigma}(p)$ and $\phi_{i\sigma}(h)$ correspond to 
the eigenfunctions of Eq.~\eqref{eq:hgw} and the index $p$ ($h$) runs
over empty (occupied) states.     

$\delta F^{\rm ph}$ contains the expansion with respect to the double-occupancy
parameters and the part of the density matrix, which commutes with the
total particle number.
This part of the RPA problem has already been studied in detail in
Refs.~\onlinecite{GOETZ1,GOETZ2}, where it was shown that
the $\delta D$ fluctuations can be eliminated 
by assuming that they adjust instantaneously to the evolution
of the density matrix (antiadiabaticity condition).

Finally, the particle-particle part of the expansion reads:
\begin{equation}
\label{espin}
\delta F^{\rm pp} = \sum_{ijkl}V_{ijkl} \kappa_{ij}^* \kappa_{kl}
\end{equation}
with $V_{ijkl}=(V_{klij})^*$ and the matrix elements of the
effective interaction are given by
\begin{eqnarray*}
V_{iiii} &=& \sum_{j} t_{ij} \left\lbrace \left(\langle c_{j\uparrow}^\dagger c_{i\uparrow}
\rangle_0 +\langle c_{i\uparrow}^\dagger c_{j\uparrow}\rangle_0\right) A_{j++}^0 A_{i++}^{''}
\right. \nonumber \\
&+& \left. \left(\langle c_{j\downarrow}^\dagger c_{i\downarrow}
\rangle_0 
+\langle c_{i\downarrow}^\dagger c_{j\downarrow}\rangle_0 \right) 
A_{j--}^0A_{i--}^{''} \right\rbrace + \frac{U}{1-n_i} \label{eq:viiii}\\
V_{iijj}&=& - t_{ij} A_i'A_j'\left\lbrace
\langle c_{j\uparrow}^\dagger c_{i\uparrow}\rangle_0
+\langle c_{j\downarrow}^\dagger c_{i\downarrow}\rangle_0\right\rbrace
\,\, \mbox{for}\;\; i \ne j \\
V_{ijjj} &=& (V_{jjij})^*= t_{ij} A_{i++}^0A_j'\hspace*{2cm}  \mbox{for}\;\; i \ne j \\
V_{ijii} &=& (V_{iiij})^*=- t_{ij} A_{j--}^0A_i' \hspace*{1.8cm} \mbox{for}\;\; i \ne j \\
V_{ijkl} &=& 0 \hspace*{5cm} \mbox{otherwise} .
\end{eqnarray*}
Here, $A_{i\tau\tau'}$ (with $\tau,\tau'=\pm$) are the matrix elements of 
Eq.~\eqref{amat} and we have defined the following abbreviations for the 
derivatives:
\begin{eqnarray} \label{z1}
A_i' &=& \left. \frac{\partial A_{i+-}}{\partial \langle J_i^-\rangle}\right|_0 =
\left. \frac{\partial A_{i-+}}{\partial  \langle J_i^+\rangle} \right|_0, \\
A_{i\tau\tau}^{''} &=& \left. \frac{\partial^2 A_{i\tau\tau}}
{\partial \langle J_i^+\rangle \partial  \langle J_i^-\rangle}\right|_0,
\label{z2}
\end{eqnarray}
where explicit expressions are given in Appendix~\ref{APA}. 
It is interesting to observe that, in contrast to the charge excitations
in the particle-hole channel, the evaluation of the pair excitations can be 
performed without any assumption on the time evolution of $\delta D$.
Only in the case of a superconducting ground state, one would have a coupling 
between (ph) and (pp) fluctuations and, therefore, the necessity to invoke the 
antiadiabaticity condition of Refs.~\onlinecite{GOETZ1,GOETZ2}
to eliminate the $\delta D$ deviations. 

Note also that in contrast to HF theory (where the expansion would be
given by $U\sum_i \delta \langle J_i^+ \rangle \delta \langle J_i^- \rangle $),
Eq.~\eqref{espin} contains correlations between distant pairs 
(i.e., $\delta \langle J_i^+\rangle \delta\langle J_j^- \rangle$)
and also processes where pairs are created and annihilated on neighboring sites
(i.e., $\delta\langle c_{i\uparrow}^\dagger c_{j\downarrow}^\dagger\rangle$). 

The remaining part of the formalism follows closely the particle-particle
RPA as developed in nuclear physics (see Refs.~\onlinecite{RING,BLAIZOT}).
We define the $\nu$-th pp-RPA eigenstate of the $N+2$ particle system by
\begin{eqnarray}\label{eq:np2}
&&|N+2,\nu\rangle =  R_\nu^\dagger |N,0\rangle, \\
&&R_\nu^\dagger = \sum_{p,p'} X^\nu_{pp'}a^\dagger_{p\uparrow}
a^\dagger_{p'\downarrow} - \sum_{h,h'}Y^\nu_{hh'}a^\dagger_{h'\uparrow}
a^\dagger_{h\downarrow},
\end{eqnarray}
where $a^\dagger_{p\sigma}$ and $a^\dagger_{h\sigma}$ create particles and
holes in the single-particle levels of the Gutzwiller Hamiltonian
Eq.~\eqref{eq:hgw}. The states $|N,\nu\rangle$ are unprojected states
in the sense of Ref.~\onlinecite{GOETZ2}, i.e., they are auxiliary
objects that have particle-particle RPA correlations   
but lack Gutzwiller correlations. This is because they result from
creating particle-particle excitations on top of $|\phi_0\rangle$ and not of
$|\Phi_G\rangle$. In the same way, the $\eta$-th eigenstate of the 
$N-2$ particle system can be represented as
\begin{eqnarray}
&&|N-2,\eta\rangle =  R_\eta^{\dagger} |N,0\rangle, \\
&&R_\eta  = \sum_{hh'} X^\eta_{hh'}a_{h\downarrow}^\dagger 
a_{h'\uparrow}^\dagger -\sum_{pp'}Y^\eta_{pp'}a_{p'\downarrow}^\dagger 
a_{p\uparrow}^\dagger,
\end{eqnarray}
where $|N,0\rangle$ is the unprojected RPA ground state of the
$N$-particle system defined by
\begin{equation}\label{eq:n0}
R_\nu |N,0\rangle = R_\eta|N,0\rangle = 0, \nonumber  
\end{equation}
and we adopt the convention that $\nu$ ($\eta$) runs over the $n_p$
($n_h$) excitations of the $N+2$ ($N-2$) particle system.
Within the pp-RPA scheme $X$ and $Y$ amplitudes can be associated with
the following unprojected matrix elements:
\begin{eqnarray*}
X^\nu_{pp'}&=& \langle N,0|a_{p'\downarrow}a_{p\uparrow}|N+2,\nu\rangle, \\
Y^\nu_{hh'} &=& -\langle N,0|a_{h\downarrow}a_{h'\uparrow}|N+2,\nu\rangle, \\
X^\eta_{hh'}&=& \langle N-2,\eta|a_{h'\uparrow}a_{h\downarrow}|N,0\rangle, \\
Y^\eta_{pp'} &=& -\langle N-2,\eta|a_{p\uparrow}a_{p'\downarrow}|N,0\rangle.
\end{eqnarray*}
From the equations of motion for the amplitudes one can derive the
following eigenvalue problem~\cite{RING,BLAIZOT}
\begin{widetext}
\begin{equation}\label{eq:motion}
\left(
\begin{array}{cc}
\lbrack \xi_{p_1\downarrow}+\xi_{p_2\uparrow}\rbrack
\delta_{p_1p_3}\delta_{p_2p_4} + V_{p_1p_2,p_3p_4} & V_{p_1p_2,h_3h_4} \\
V_{h_1h_2,p_3p_4}^* & -\lbrack
\xi_{h_1\downarrow}+\xi_{h_2\uparrow}\rbrack
\delta_{h_1h_3}\delta_{h_2h_4} + V_{h_1h_2,h_3h_4} \end{array} \right)
\left(\begin{array}{c}
W_{p_3p_4} \\ Z_{h_3h_4} \end{array}\right) =  
\left(\begin{array}{cc}
I_{n_p} & 0 \\
0 & -I_{n_h} \end{array}\right)
\left(\begin{array}{c}
W_{p_1p_2} \\ Z_{h_1h_2} \end{array}\right) \Omega^\pm,
\end{equation}
\end{widetext}
with $I_{n}$ the $n\times n$ identity matrix. The matrix elements of
the potential $V$ can be derived from Eq.~\eqref{espin} by
transforming to the GA representation with the help of Eq.~\eqref{eq:trafo}.

Equations~\eqref{eq:motion} yield $n_p+n_h$ eigenvectors which can be
normalized as $|W|^2-|Z|^2=\pm 1$. The sign of the norm allows one to
distinguish the $n_p$ addition eigenvectors (positive norm) from the 
$n_h$ removal eigenvectors (negative norm).~\cite{BLAIZOT}

In the following, we will denote by $E_\nu(N)$ and $E_\eta(N)$ the $N$-particle
energies of the Hamiltonian Eq.~\eqref{HM} when the $\mu$ term is absent 
and write the $\mu$ contribution explicitly, so that the eigenvalues of 
Eq.~\eqref{HM} are given by $E_\eta(N)-\mu N$. The eigenvalues and 
eigenvectors can be identified with the excitation energies and the amplitudes 
in the following form:
\begin{eqnarray}
\Omega^+_{\nu} &=& E_\nu(N+2) - E_0(N)-2\mu, \label{en2add}\\ 
W^\nu_{pp'}&=&X^\nu_{pp'}, \nonumber\\
Z^\nu_{hh'}&=&Y^\nu_{hh'}, \nonumber
\end{eqnarray}
for two-particle addition and
\begin{eqnarray}
\Omega^-_{\eta} &=& E_0(N) - E_\eta(N-2)-2\mu, \label{en2min}\\
 W^\eta_{pp'}&=&Y^\eta_{pp'}, \nonumber\\
Z^\eta_{hh'}&=&X^\eta_{hh'}, \nonumber 
\end{eqnarray}
for two-particle removal.

Stability requires that $\Omega^+_{\nu}>0>\Omega^-_{\eta}$ thus
if there exists a chemical potential $\mu$ such that these two conditions 
are satisfied the system is stable, otherwise a pairing instability arises. 

In a finite system, the allowed range of $\mu$ shrinks to zero as an 
instability is approached. From Eqs.~\eqref{en2add} and~\eqref{en2min} one 
sees that stability requires $E_0(N+2) - E_0(N)> E_0(N) - E_0(N-2)$. This
coincides with the stability against a transfer of a pair of particles 
from one cluster to another one in an ensemble. Notice that it does
not necessarily coincide with the stability against single particle
transfers. In the thermodynamic limit both conditions coincide with the well 
known stability condition that the compressibility must be positive. 

We are now in the position to evaluate the pairing correlation function 
within the TDGA. We are interested in on-site s-wave (s) pairing and intersite 
pairing:
\begin{eqnarray}
\Delta^{s}_i &=& c_{i\downarrow}c_{i\uparrow}, \label{eq:son}\\
\Delta^{\delta}_i &=&\frac1{\sqrt2}
(c_{i+\delta\downarrow}c_{i\uparrow}+c_{i\downarrow}c_{i+\delta\uparrow}),
\label{isite} 
\end{eqnarray}
where $\delta=x,y$ and $i+\delta$ is a shorthand for the first
neighbor of site $i$ in the $\delta$ direction. 
 
The dynamical pairing correlations can be computed from:
\begin{widetext}
\begin{eqnarray}\label{eq:pcorr}
P_{ij}^{\alpha\beta}(\omega)&=& -\frac{\imath}{N_s} \int_{-\infty}^{\infty}dt 
\mbox{e}^{\imath\omega t}\langle {\cal T} \Delta_i^\alpha(t)
[\Delta_j^\beta(0)]^\dagger\rangle \\
&=&
\frac{1}{N_s}\sum_\nu \frac{\langle N,0|\bar\Delta_i^\alpha|
N+2,\nu\rangle\langle N+2,\nu|(\bar\Delta_j^\beta)^\dagger|N,0\rangle}
{\omega - \Omega^+_\nu + \imath0^+}
- \frac{1}{N_s}\sum_\eta \frac{\langle N,0|(\bar\Delta_j^\beta)^\dagger|
N-2,\eta\rangle\langle N-2,\eta|\bar\Delta_i^\alpha|N,0\rangle}
{\omega - \Omega^-_\eta - \imath0^+}, \nonumber
\end{eqnarray}
\end{widetext}
where $N_s$ denotes the number of sites of the system and
$\alpha=s,x,y$. The average in
the first line is done on the Gutzwiller projected RPA state. The latter is
our best estimate for the true ground state of the system. 
The matrix elements in the second line are done on the unprojected
states. In the spirit of the GA, Gutzwiller projections are effectively taken 
into account by a renormalization of the operators. As usual local operators 
do not get renormalized, i.e., $\bar\Delta_i^s\equiv \Delta_i^s$, whereas 
non-local operators acquire a renormalization through the $z$ factors: 
\begin{equation}\label{exswave}
\bar\Delta_i^{\delta}\equiv \frac1{\sqrt2}
(z_{i+\delta\downarrow}z_{i\uparrow}c_{i+\delta\downarrow}c_{i\uparrow}+
z_{i\downarrow}z_{i+\delta\uparrow}c_{i\downarrow}c_{i+\delta\uparrow}).
\end{equation}
This is similar to the
renormalization of the GA current operator in the computation of the optical
conductivity.~\cite{GOETZ2} 
The validity of this prescription can be checked using sum rules as
discussed below. Obviously in the HF theory based on BLA, bare 
operators are used in Eq.~\eqref{eq:pcorr}.

The matrix elements of the pairing operators can be obtained from the 
amplitudes $X$ and $Y$ using the transformation Eq.~\eqref{eq:trafo}. 
In Eq.~\eqref{eq:pcorr} the first and second term represent the correlations 
in case of two-particle addition and removal, respectively.

In the case of a separable potential  
$V_{k_1k_2,k_3k_4}=v_{k_1k_2}v_{k_3k_4}^*$ the pp-RPA equations can be
easily solved. In particular for the case of a paramagnetic solution analyzed 
in Ref.~\onlinecite{sei08} the  TDGA interaction kernel Eq.~\eqref{espin} 
only involves local pairs 
\begin{equation}\label{espin2}
\delta F^{\rm pp} = V \sum_{i}\delta\langle c_{i\uparrow}^\dagger
c_{i\downarrow}^\dagger\rangle \delta\langle c_{i\downarrow} c_{i\uparrow}\rangle
\end{equation}
with
\begin{equation}\label{eq:veff}
V=\frac{U-2\Sigma_\sigma}{1-n},
\end{equation}
where $n\equiv N/N_s$ and $\Sigma_\sigma$ is defined in 
Eq.~\eqref{eq:lagrange}. In this case the on-site response can be found in 
momentum space:
\begin{eqnarray}
P^{\alpha\alpha}_{ii}(\omega)&=&\frac{1}{N_s}\sum_{\bf q} P_{\alpha\alpha}({\bf q},\omega) \\
P_{\alpha\alpha}({\bf q},\omega)&=& -\frac{\imath}{N_s} 
\int_{-\infty}^{\infty}dt \mbox{e}^{\imath\omega t}
\langle {\cal T} \Delta^\alpha_{\bf q}(t)
[\Delta^\alpha_{\bf q}(0)]^\dagger \rangle \nonumber.
\end{eqnarray}
with ($\alpha=s$)
\begin{eqnarray}
\Delta^\alpha_{\bf q} &=& \frac{1}{N_s} \sum_j\Delta^\alpha_i e^{i {\bf q}.{\bf R}_j} 
\end{eqnarray}
and is given by an effective ladder type equation:
\begin{eqnarray}
P_{ss}({\bf q},\omega) &=& \frac{P^0_{ss}({\bf q},\omega)}{1-V P^0_{ss}({\bf q},\omega)}
\label{eq:prpa} \\
P^0_{ss}({\bf q},\omega)&=& \frac{1}{N_s} \sum_{\bf k} \frac{1-f(\xi_{\bf k})  
- f(\xi_{{\bf k+q}})}{\omega - \xi_{\bf k} -
\xi_{{\bf k+q}}+ \imath \eta_{{\bf k},{\bf k+q}}}\label{chi0}.
\end{eqnarray}
with $f(\xi_{\bf k})$ the Fermi distribution function,
and $\eta_{{\bf k},{\bf k'}}\equiv 0^+ {\rm sign}(\xi_{\bf k}+\xi_{{\bf k}'})$.
The single particle energies are given by 
$\xi_{\bf k}=\varepsilon_{\bf k}-\mu$ where $\varepsilon_{\bf k}\equiv z_0^2
e_{\bf k}+\Sigma_G$ is the GA dispersion relation and 
$e_{\bf k}=\sum_jt_{ij} e^{-i{\bf k}.({\bf R}_j-{\bf R}_i)}$ is the
bare one and $z_0$ is the hopping renormalization factor
Eq.~\eqref{qfac} evaluated at the saddle point.  

We are interested also in the fluctuation response for intersite
bond singlets with s-wave symmetry
($\alpha=+$) or d-wave ($\alpha=-$) symmetry:

\begin{equation}
\Delta^\alpha_i = \frac{1}{2}  \label{eq:deltaa}
\left( \Delta_{i}^x + \Delta_{i}^{-x} + \alpha \Delta_{i}^{y} + \alpha
      \Delta_{i}^{-y} \right)\nonumber
\end{equation}
In this case the ladder series takes the following matrix form: 
\begin{equation}\label{eq:pis}
\underline{\underline{P}}({\bf q},\omega) = \underline{\underline{P}}^0({\bf q},\omega) 
+ \underline{\underline{P}}^0({\bf q},\omega) \underline{\underline{\Gamma}}\,\,
\underline{\underline{P}}({\bf q},\omega)
\end{equation}
with
\begin{displaymath}
\underline{\underline{P}}({\bf q},\omega)= \left( \begin{array}{cc} 
P_{ss}({\bf q},\omega) & P_{s\alpha}({\bf q},\omega) \\ 
P_{\alpha s}({\bf q},\omega) & P_{\alpha\alpha}({\bf q},\omega)
\end{array} \right) \,\, ; \,\,
\underline{\underline{\Gamma}}= \left( \begin{array}{cc} 
V & 0 \\ 
0 & 0
\end{array} \right).
\end{displaymath}
Thus
\begin{equation}
  \label{eq:pint}
  P_{\alpha\alpha}({\bf q},\omega)=P_{\alpha\alpha}^0({\bf
  q},\omega)+\frac{P_{\alpha s}^0({\bf q},\omega)V P_{s \alpha }^0({\bf q},\omega)}{1-V P_{ss}^0({\bf q},\omega)}
\end{equation}
and the poles of $P_{\alpha\alpha}({\bf q},\omega)$ are determined
by the local pair correlation ($1-V P_{ss}^0({\bf q},\omega)=0$) and
the bare correlation. 

Eqs.~\eqref{eq:prpa}-\eqref{eq:pint} are also
valid in the BLA approach, replacing the Gutzwiller local self-energy 
with the HF one ($\Sigma_\sigma = U n/2$, $z_0=1$) and taking  $V=U$.

The exact on-site s-wave spectral density satisfies  the following sum rules:  
\begin{eqnarray}
  -\frac1\pi \int_{0}^{\infty} d\omega {\rm Im}\,
  P^{ss}_{ii}(\omega)&=&1-n+\langle n_{i\uparrow}n_{i\downarrow}\rangle
\label{eq:intp1}\\
  -\frac1\pi \int_{-\infty}^{0} d\omega {\rm Im}\,P^{ss}_{ii}(\omega)&=&-\langle n_{i\uparrow}n_{i\downarrow}\rangle.
\label{eq:intp2}
\end{eqnarray}
In our case these sum rules allow us to compute RPA corrections to the double
occupancy and they will be used below to evaluate RPA corrections to the
GA ground state energy. Notice that the total spectral weight is equal
to $1-n$. 

In addition the following energy weighted sum rule is satisfied
\begin{eqnarray}
M_1^{ss}(\infty) &=& \frac2N_s\langle T\rangle_{GA}+(2\mu-U) (1- n) 
\label{eq:sum}\\
M_1^{ss}(\omega) &=&
\frac{1}{\pi N_s}\sum_{ij}\int_{-\infty}^{\omega}\!\!\!\!d\bar{\omega}\,
\bar{\omega}{\rm Im} P_{ij}^{ss}(\bar{\omega}) .
\end{eqnarray}
with $T$ the kinetic energy. This sum rule is satisfied exactly within the 
present pp-RPA in the sense that the right and the left hand side are equal,
provided the kinetic expectation value on the r.h.s. is computed at the GA 
level. Thus in contrast with Eqs.~\eqref{eq:intp1} and~\eqref{eq:intp2} 
this sum rule provides no new information but is useful to perform a 
consistency check. The same prescription is valid in the conventional HF plus 
pp-RPA approach, where the r.h.s. expectation values have to be computed at HF 
level.~\cite{THOULESS} As shown in Appendix~\ref{APB} the consistent 
renormalization of intersite operators can be checked from an analogous sum 
rule.

\section{Results}\label{section:III}

In this section we present results for pair correlations in the repulsive 
Hubbard model within the framework developed above. Since the RPA-type
scheme used in the TDGA differs in some regard from the approaches usually
invoked in solid-state theory, we first illustrate the method for
a two-site model which also can be solved exactly. Due to the
mean-field character of the present approximations the method is
expected to improve with dimensionality so this zero-dimensional example
represents the worst case and allows us to give a first check of the
potentialities and limitations of the method.
Finally, we compare our method with exact results on small clusters and
demonstrate its superior performance as compared to the BLA.
    
\subsection{Two-site model}

In order to illustrate the formalism we consider a two-site model with two 
particles having up and down spins. The ground-state wave function reads as
\begin{equation}
|2,0 \rangle = \alpha |S\rangle + \beta |D\rangle,
\end{equation}
where
\begin{eqnarray}
|S\rangle &=& \frac{1}{\sqrt{2}}\left(|1_\uparrow 2_\downarrow \rangle 
- |1_\downarrow 2_\uparrow \rangle \right),\\
|D\rangle &=& \frac{1}{\sqrt{2}}\left(|1_\uparrow 1_\downarrow \rangle 
+ |2_\uparrow 2_\downarrow \rangle \right),
\end{eqnarray}
and the corresponding amplitudes and the ground-state energy $E_0(2)$
are given by
\begin{eqnarray}
\alpha &=& \frac{2t}{\sqrt{\omega_0^2 + 4t^2}}, \\
\beta &=& \frac{-\omega_0}{\sqrt{\omega_0^2 + 4t^2}}, \\
E_0(2)&=&\omega_0\equiv \frac{U}{2}\left\lbrack 1-\sqrt{1+\frac{16t^2}{U^2}}
\right\rbrack
\end{eqnarray}
For four and zero particles there is only one state with energy
$E(4)=2U$ and $E(0)=0$, respectively. 
The energy differences between two-particle addition 
(removal) states and the ground state are 
\begin{eqnarray*}
\Omega^+=E(2+2)-E_0(2)-2\mu &=&  \frac{U}{2}
+\frac{1}{2}\sqrt{U^2+16t^2}, \\
\Omega^-= E_0(2) - E(2-2)-2\mu  &=& -\Omega^+,
\end{eqnarray*}
and the chemical potential is taken as $\mu=U/2$ which is the exact
value at half filling at an infinitesimal temperature. 
The corresponding matrix elements for local ($s$) and intersite ($x$) pairing
operators
\begin{eqnarray} 
\Delta_s &=& \frac1{\sqrt{2}}(c_{1\downarrow}c_{1\uparrow} 
+ c_{2\downarrow}c_{2\uparrow}) \label{local2s}, \\
\Delta_x &=& \frac1{\sqrt{2}}(c_{1\downarrow}c_{2\uparrow} 
- c_{1\uparrow}c_{2\downarrow}), \label{inter2s}
\end{eqnarray}
read as
\begin{eqnarray}
\langle 2,0|\Delta_s|2+2\rangle &=& \langle 2,0|\Delta_s^\dagger|2-2\rangle
= \beta, \\
\langle 2,0|\Delta_x|2+2\rangle &=& \langle 2,0|\Delta_x^\dagger|2-2\rangle
= \alpha.
\end{eqnarray}
Notice that there is only one state with $2\pm2$ particles so we dropped the 
excitation index. One can check that the sum rules of
Eq.~\eqref{eq:intp1}-\eqref{eq:sum} are satisfied. For example the
double occupancy is given by $\beta^2/2$ and the first moment sum rule   
\begin{eqnarray*}
&-&\frac{2}{N_s}\langle T\rangle -(2\mu-U) (1-n) \\
&=&\Omega^{+}|\langle 2,0|\Delta_s|2+2\rangle|^2
- \Omega^{-}|\langle 2,0|\Delta_s^\dagger|2-2\rangle|^2 \\
&=& 2(U - \omega_0)\beta^2 = -\langle T\rangle
\end{eqnarray*}
is also fulfilled, as it should.

Consider now the same model within the TDGA. On the GA level, one finds two 
single particle states for each spin direction which at half filing can be 
put as:
\begin{eqnarray}
\xi_{h\sigma} &=& -t(1-u^2) +\Sigma_\sigma -\mu, \\
\xi_{p\sigma} &=& t(1-u^2) +\Sigma_\sigma  -\mu.
\end{eqnarray}
where the h-states is occupied with a spin-up and spin-down particle and we 
have defined $u\equiv U/U_{BR}$ with $U_{BR}=8t$. For $U<U_{BR}$ there is a 
paramagnetic solution which becomes insulating at the 
Brinkmann-Rice transition point $U_{BR}$.~\cite{bri70}
For $u > \sqrt{2}-1$ the more stable solution is an antiferromagnetic  
broken symmetry solution which does not have a Brinkmann-Rice 
transition point. 

For the paramagnetic solution one has spin-independent Lagrange multipliers 
which from the Kotliar-Ruckenstain (or Gebhard's) scheme are obtained as
\begin{equation}
\Sigma_\sigma \equiv \Sigma_0 = \mu = \frac{U}{2}.
\end{equation}
Since there is only one two-particle addition and removal state,
we have to diagonalize the following $2\times 2$ pair fluctuation RPA problem
\begin{widetext}
\begin{equation}
\left(\begin{array}{cc}
2t(1-u^2)+V/2 & V/2  \\
V/2 & 2t(1-u^2)+V/2
\end{array}\right) \left(\begin{array}{c}
X \\ Y \end{array}\right) 
= \Omega \left(\begin{array}{cc}
1 & 0 \\
0 & -1
\end{array}\right)\left(\begin{array}{c}
X \\ Y \end{array}\right) \label{rpa2x2}
\end{equation}
\end{widetext}
and the interaction $V$ corresponds to the local part in the expansion
Eq.~\eqref{espin} (note that the first derivatives of $A_i^0$ vanish for
a paramagnetic solution)
\begin{equation}
V=-2t(1-u^2) A'' + \frac{U}{1-n} = 4 t u (2-u) \frac{1+u}{1-u}
\end{equation}
where $A''$ is defined in Appendix~\ref{APA}. Here, the diverging part,
proportional to $1/(1-n)$, is canceled by an analogous contribution in the 
first term. The same result can be obtained from Eq.~\eqref{eq:veff} by taking 
the limit $n\rightarrow 1$,~\cite{sei08} and coincides in modulus with the
effective interaction at half filling in the particle-hole
case.~\cite{VOLLHARDT,GOETZ1,GOETZ2} This is consistent with the fact that the
attractive-repulsive transformation converts particle-particle
fluctuations in particle hole-fluctuations with equal interaction but
sign reversed.~\cite{mic90} 

Diagonalization of Eq.~\eqref{rpa2x2} yields the two eigenvalues
\begin{equation}
\Omega^{\pm} = \pm 2t(1-u^2)\sqrt{1+\frac{V}{2t(1-u^2)}}.
\end{equation}
For the eigenvectors, the following relations hold
\begin{eqnarray}
(X^\pm + Y^\pm)^2 &=& \pm \frac{2t(1-u^2)}{\Omega^\pm},\\
(X^\pm)^2 - (Y^\pm)^2 &=& \pm 1, 
\end{eqnarray} 
from which one can compute the amplitudes for the local and intersite pairing 
correlations between {\em unprojected} states: 
\begin{eqnarray}
\langle 2,0|\Delta_s|2\pm 2\rangle &=& \frac{1}{\sqrt{2}}(X^\pm + Y^\pm),
\label{eq:matl} \\
\langle 2,0|\Delta_x|2\pm 2\rangle &=& -\frac{1}{\sqrt{2}}(X^\pm - Y^\pm).
\label{eq:matis}
\end{eqnarray}
Finally, the expectation values between Gutzwiller {\em projected} states are 
the same as for unprojected states in the case of $\Delta_s$ and should be 
renormalized in the intersite case, i.e., $\bar\Delta_x=(1-u^2)\Delta_x$. 
See Appendix~\ref{APB} for a consistency check of this prescription.

For the local pairing operator we find that the
first moment sum rule Eq.~\eqref{eq:sum}
\begin{eqnarray}
&&\Omega^+ |\langle 2,0|\Delta_s|2+2\rangle|^2
- \Omega^- |\langle 2,0|\Delta_s^\dagger|2-2\rangle|^2  \\
&=& 2t(1-u^2) = -\frac{2}{N} \langle T\rangle_{GA} -2\mu(1-n) +U(1-n) \nonumber\end{eqnarray}
is satisfied (note that $n=1$) in the TDGA, as anticipated. 

\begin{figure}[htb]
\includegraphics[width=8.5cm,clip=true]{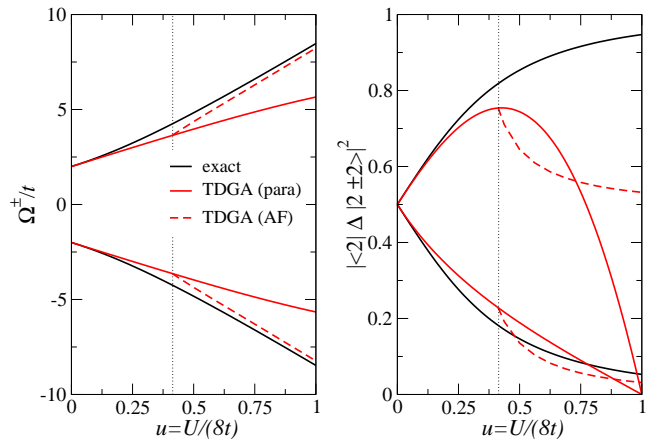}
\caption{
(color online) Left panel: Excitation energies for two-particle addition and removal computed 
for the two-site model with two particles. Right panel: Matrix element of the 
local (lower curves) and intersite (upper curves) pairing operator. 
The underlying GA saddle-points are paramagnetic (p) (solid lines) and 
antiferromagnetic (dashed lines). The critical value $u=\sqrt{2}-1$ of the
corresponding transition is indicated by the vertical dotted line.}
\label{fig2s1}
\end{figure}

The TDGA two-particle addition and removal energies are displayed in 
Fig.~\ref{fig2s1} and compared with the exact ones. Solutions obtained in the
paramagnetic regime are shown with solid lines. 
Remarkably, due to a cancellation of $(1-u)$ terms the Brinkmann-Rice 
transition does not reflect in the excitation energies of the paramagnetic 
phase, which become soft at a much larger value of the repulsion
($u=1+\sqrt2$). Instead, the transition shows up in the TDGA pairing
correlations (right panel of Fig.~\ref{fig2s1}). Indeed,
the Gutzwiller renormalization factors $z_{i\sigma}$ drive the matrix
elements of the intersite pairing operator to zero at the Brinkmann-Rice
transition point. It is interesting that neglecting $z_{i\sigma}$
in the pairing operator, the suppression is replaced by a divergence so that 
the $z_{i\sigma}$ cancel the unphysical divergence but overcorrect
it. This problem is partially cured in the broken symmetry state (dashed lines)
but for $U\to\infty$ the matrix element tends to 1/2 whereas the 
exact result approaches $|\langle 2-2 | \Delta_x| 2,0 \rangle|^2 \to
1$ since the ground state can be written as  
$\Delta_x^{\dagger}  | 2-2 \rangle $ in this limit. Instead, the
ground state with broken symmetry has only one configuration of the 
two that generate the singlet ground state which explains
the 1/2 factor. For the local pairing operator the behavior is better.

We notice that the TDGA excitations energies are in quite
good agreement with the exact values, especially when one allows for the
broken symmetry solutions.
 
\begin{figure}[htb]
\includegraphics[width=8.5cm,clip=true]{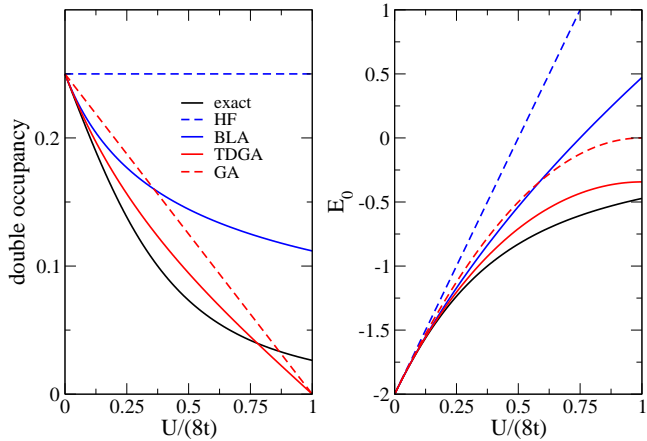}
\caption{(color online) 
Double occupancy (left panel) and ground-state energy (right panel) 
of the two-site Hubbard model computed within HF, BLA, TDGA, and exact 
diagonalization. Only paramagnetic ground states are considered.}
\label{fig2s2}
\end{figure}

Finally we can use Eq.~\eqref{eq:intp1} to compute 
the pair fluctuation derived  double occupancy:  
\begin{equation}\label{eq:dble}
D^{RPA}=\langle n_{i\uparrow}n_{i\downarrow}\rangle
= \langle 2,0|c_{i\downarrow}c_{i\uparrow}|2+2\rangle
\langle 2+2|c^\dagger_{i\uparrow}c^\dagger_{i\downarrow}|2,0\rangle.
\end{equation}
Fig.~\ref{fig2s2} (left panel) shows the double occupancy in different
approximations. HF completely neglects correlations and hence the
double occupancy is independent of $U$. The BLA based correction drives 
the approximation much closer to the exact result at
small $U$ but strongly underestimates the correction at large $U$. 
In the GA, correlations are taken into account already
at the static level and the double occupancy is strongly suppressed as
a function of $U$. Thus the pp-TDGA correction is small and brings the
double occupancy close to the exact one in a much larger range of
interaction. The fact that the  pp-TDGA is close to the GA double
occupancy indicates that the theory is nearly self-consistent. In 
Ref.~\onlinecite{sei08} it was shown that this feature is enhanced in
two-dimensions pointing to an improvement of the performance as the
dimensionality is increased. In contrast the BLA is clearly quite far
from being self-consistent in this sense.

For large $U$, the exact double occupancy is of order
$UJ \sim t^2/U^2$ due to the same charge fluctuations that build
the double exchange interaction $J$.~\cite{lor05} 
Since the GA and the TDGA results in Fig.~\ref{fig2s2} are for paramagnetic
solutions the double occupancy vanishes at the Brinkman-Rice
point. This makes clear that the scale $J$ is not present in the
paramagnetic GA or TDGA.

The above results allow us to compute the pp-RPA ground state energy
using the coupling constant integration trick~\cite{mahan,fetter} 
\begin{equation}\label{eq:coupl}
E_0^{RPA}=-2t+U N_s \int_0^U dU' D^{RPA}(U').
\end{equation}
Here the first term is the ground state energy for $U=0$. 
Restricting on the paramagnetic solutions, we find for the TDGA and BLA
\begin{eqnarray}
E_0^{TDGA}&=& -2t + 4t(\sqrt{1+2u-u^2}-1),\\
E_0^{BLA}&=& -2t + 2t(\sqrt{1+U/2}-1),
\end{eqnarray}
which both yield the exact result $\omega_0$ up to second order in $U/t$.

Due to the more accurate estimate for the double occupancy
(see Fig.~\ref{fig2s2}a) it turns out that the TDGA gives a 
significantly better approximation for the ground-state energy 
(see Fig.~\ref{fig2s2}b) than the BLA over the whole range of $U$, before 
the onset of the Brinkmann-Rice transition. 
This should be compared with an analogous calculation in the particle-hole
channel in Ref.~\onlinecite{GOETZ2}. In the latter case one obtains a
singularity in $E_0$ at $u=8(\sqrt{2}-1)$ due to the onset of
the antiferromagnetic ground state. Since there is no instability in the 
particle-particle channel, such problems do not arise in the present case. 
Thus it seems convenient in general to compute RPA corrections to the energy 
in a channel free from instabilities. 

\subsection{Comparison with exact diagonalization}
 
In this section, we study the dynamical pairing correlations within
the TDGA on small clusters and compare our results with the BLA approach and
exact diagonalizations.

We start by computing the long-distance pairing correlations
$\langle\Delta_i^{\delta'}  (\Delta_j^\delta)^\dagger\rangle$ for bond 
singlets. Within the pp-RPA these correlations are obtained integrating the 
addition part of the dynamical spectral function:
$$\langle \Delta_i^{\delta'} (\Delta_j^{\delta})^\dagger\rangle=
\frac1\pi \int^{\infty}_{0} d\omega {\rm Im}
P_{ji}^{\delta'\delta}(\omega).$$
Thus the comparison with the exact diagonalization results provides a
stringent test of the total spectral weight. We remind the reader
that in the TDGA the singlet operator is renormalized by the  
$z_{i\sigma}$ factors. 

\begin{figure}[htb]
\includegraphics[width=8cm,clip=true]{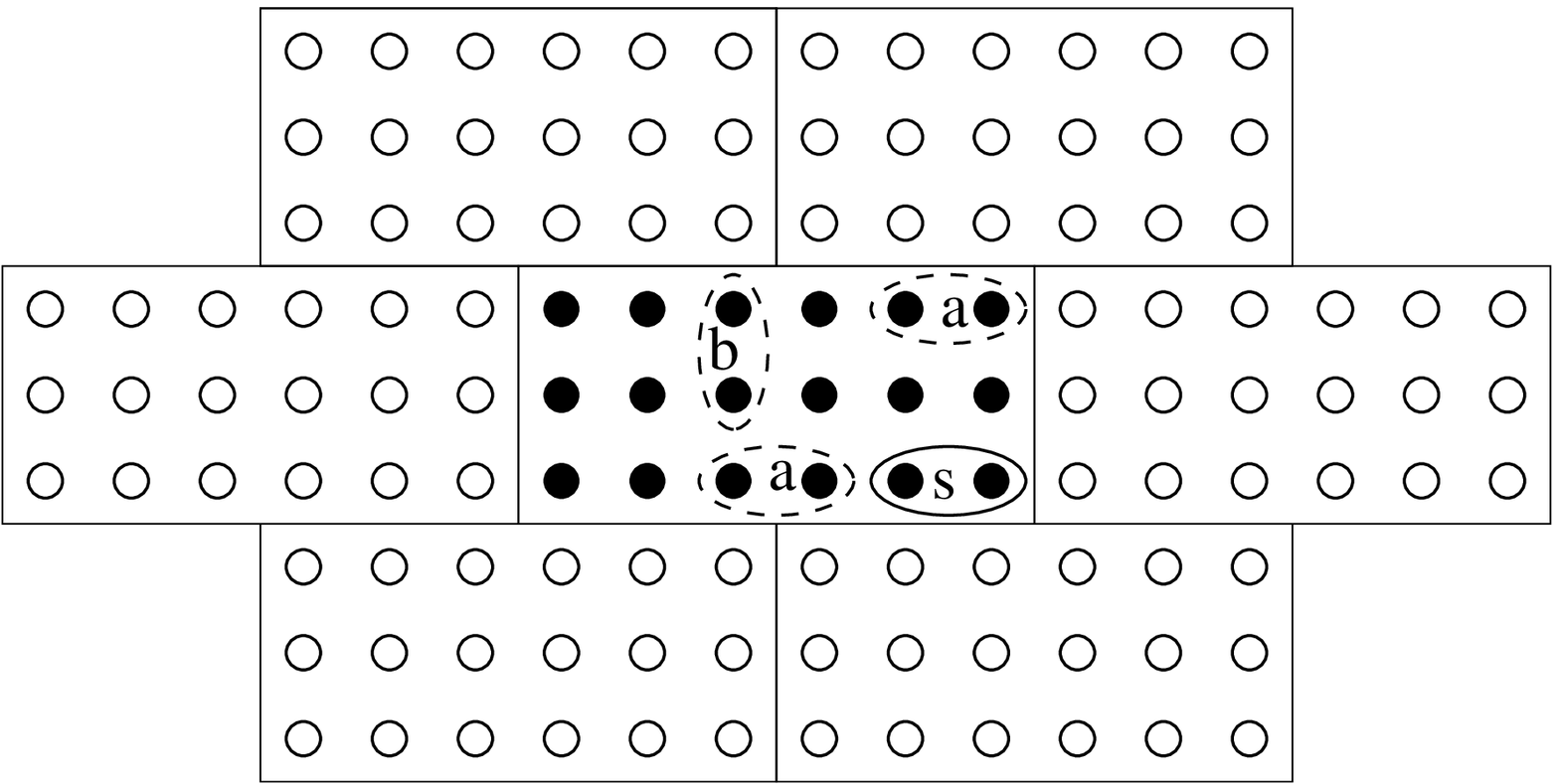}

\includegraphics[width=8cm,clip=true]{fig3b.eps}
\caption{(color online) 
Top panel: Sketch of the 18-site cluster and corresponding boundary conditions.
Bottom panels: Bond singlet pairing correlations on the 18-site lattice
for $U/t=4$ and $10$. The bars at the TDGA symbols point to the bare GA value 
of the singlet correlations. The separation between bond singlets
is defined as the shortest distance between their centers. Taking the 
singlet `s' as reference, it should be noted that there exist two parallel 
singlet configurations `a' with distance $R_s-R_a=2$. The singlets `b' and `s' 
form a perpendicular configuration.}
\label{figcorr}
\end{figure}

In Fig.~\ref{figcorr} we show results for 10 particles on a cluster with 18 
sites, tilted by $45^\circ$, and, therefore, having all the spatial symmetries
of the infinite lattice. A particular representation of this cluster, with its
boundary conditions, is shown in the upper panel of Fig.~\ref{figcorr}. 
In this case, the GA ground state is paramagnetic. We concentrate on 
non-overlapping singlets and the distance between them is measured from their 
centers. 

The two singlet operators can form a perpendicular configuration, like
`s-b' in the upper panel of Fig.~\ref{figcorr}, or a parallel configuration, 
like `s-a' in the same figure. Notice that for $R_{ij}=R_i-R_j=2$ 
there are two points in the lower panel corresponding to two parallel 
configurations in which one of the  singlets is displaced either along 
the x- or along the y-direction (labeled `a' in the upper panel of 
Fig.~\ref{figcorr}). 

In Fig.~\ref{figcorr}, the vertical bars point to the GA value of 
the singlet correlations, i.e., the decoupled result of 
$\langle \Delta_i^{\delta'}(\Delta_j^\delta)^\dagger\rangle_{GA}$ but still 
renormalized with the $z_{i\sigma}$ factors. In the same spirit of
Ref.~\onlinecite{white89} the length and orientation of the bars reflects 
the `vertex contribution'. This quantity measures the correlation induced
interaction between two singlets which is attractive when the TDGA value
is larger than the one computed within the bare GA.
For nearby singlets we observe excellent agreement between the exact
diagonalization result and the TDGA for both $U/t=4$ and $10$
(see lower panels of Fig.~\ref{figcorr}). In this case we also observe
an effective attractive interaction. In general, with increasing distance TDGA 
overestimates the exact correlations, and the difference becomes 
more pronounced for larger $U/t$. This behavior can be expected due to the fact
that the Gutzwiller method is based on a local projector which neglects 
intersite correlations.
One can therefore anticipate that the incorporation of intersite projections
(as in Jastrow-type wave functions~\cite{jast55,capello}) into the TDGA would 
lead to an improvement of the long-distance pair correlations. Nevertheless, 
it turns out that the TDGA yields a rather good description of long-distance 
pair correlations (especially for moderate values of $U/t$) as compared to 
exact diagonalizations. 

In order to compare dynamical properties of the TDGA with BLA and
exact results, we compute the d-wave and s-wave correlations 
of Sec.~\ref{sec2b}. The comparison of the results in different approximations
can be done by aligning the chemical potentials (as in the last figure of 
Ref.~\onlinecite{sei08}) or aligning the absolute energies. In the
following, we adopt the last procedure by eliminating the chemical
potential from the response function. We define $\omega'=\omega+2\mu$
so that the poles in $P^{\alpha\alpha}_{ii}(\omega')$ occur at
$\omega'=E_\nu(N+2)-E_0(N)$ and $\omega'=E_\eta(N-2)-E_0(N)$. 

\begin{figure}[htb]
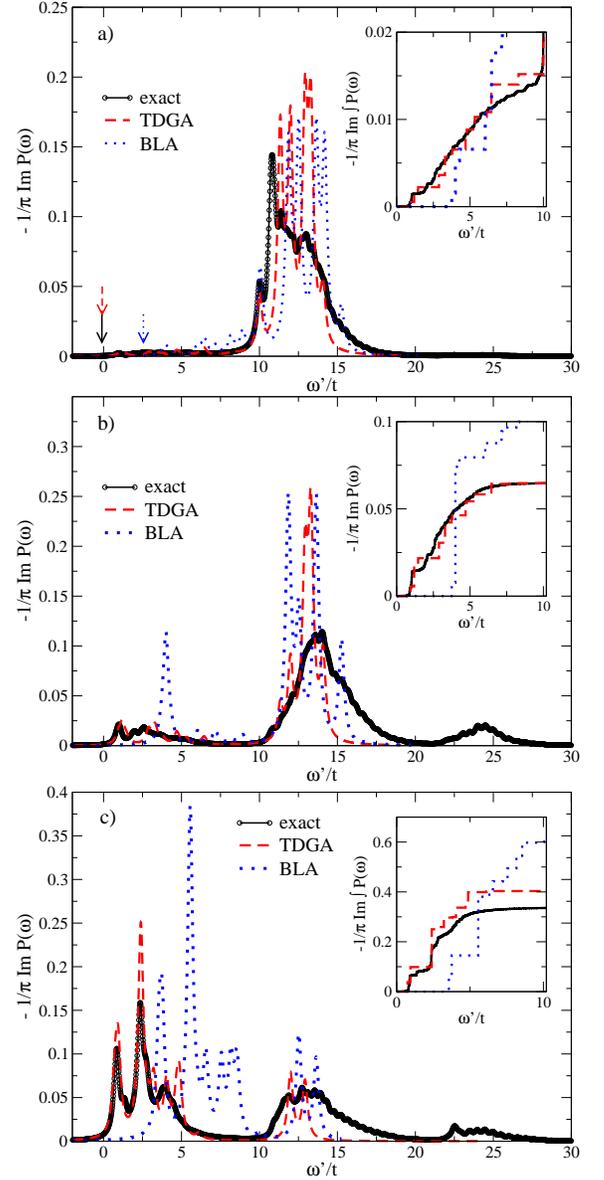

\includegraphics[width=7.5cm,clip=true]{fig4a.eps}
\includegraphics[width=7.5cm,clip=true]{fig4b.eps}
\includegraphics[width=7.5cm,clip=true]{fig4c.eps}
\caption{(color online) 
Imaginary part of the (two-particle addition) pairing correlation 
function of Eq.~\eqref{eq:pcorr} for on-site pairing (a), extended s-wave (b) 
and d-wave (c) symmetry. Results are for 10 particles on a 18-sites cluster 
and $U/t=10$ for which the underlying mean-field solution in BLA and TDGA is 
paramagnetic. The delta-like excitations are convoluted by lorentzians with 
width $\epsilon = 0.2t$. The arrows in the upper panel point to the chemical
potentials obtained within the various methods.
The insets detail the frequency evolution of the weight at small energies.}
\label{figresu10}
\end{figure}

\begin{figure}[htb]
\includegraphics[width=7.5cm,clip=true]{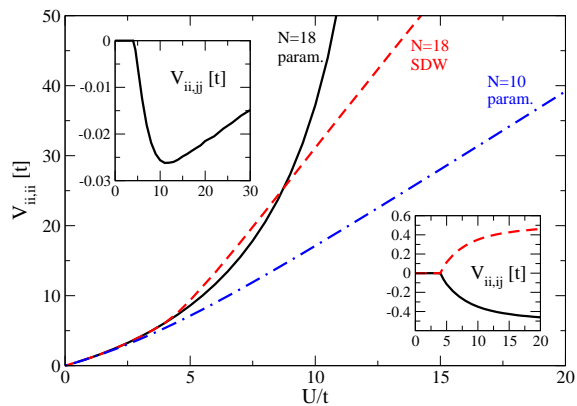}
\caption{(color online) 
Main panel: Local part of the effective interaction $V_{ii,ii}$  
(see Eq.~\eqref{espin}) for the 18-site cluster. The solid and dashed lines 
refer to the half-filled cluster with paramagnetic and spin-density wave 
ground state, respectively. The dashed dotted line is for the 10-particle 
system with a paramagnetic ground state. 
Upper-left inset: Effective interaction between pairs on nearest-neighbor 
sites $V_{ii,jj}$. Lower-right inset: Effective interaction $V_{ii,ij}$ 
between a local pair on site $R_i$ and an intersite pair on the bond defined 
by $R_i$ and $R_j$. $V_{ii,ij} > (<) 0$ for $S_i^z < (>) 0$ 
(solid and dashed lines respectively).}
\label{figueff}
\end{figure}

In Fig.~\ref{figresu10}, we show the addition spectra for 
$P^{\alpha\alpha}_{ii}(\omega')$ evaluated for a 18-site cluster with 10 
particles and $U/t=10$. This corresponds to a closed shell configuration 
where the HF and GA approach yield paramagnetic solutions. 
The on-site s-wave case $P^{ss}_{ii}(\omega')$ has already been
analyzed~\cite{sei08} for a smaller cluster and it is shown in 
Fig.~\ref{figresu10} for 18 sites a as reference since, as explained in 
Sec.~\ref{sec2b}, it also determines the overall features of the intersite 
pairing correlations. Fig.~\ref{figueff} (main panel, dashed dotted line) 
displays the effective interaction $V$ for the present case as a function 
of $U/t$.

In Ref.~\onlinecite{sei08} we have shown that the on-site pair excitations for 
large $U$ in both BLA and TDGA are dominated by an antibound which lower 
band edge is at $\omega'=U$. The band states at energies $1t\sim 5t$ have 
very small spectral weight. In the top panel of Fig. \ref{figresu10} the
full black arrow indicates the value of $2\mu$ in the exact case with
$\mu=(\mu^++\mu^-)/2$ and $\mu^+=E(N+1)-E(N)$, $\mu^-=E(N)-E(N-1)$ . 
The position of $2\mu$ in the GA (red dashed arrow) is very close
to the exact one whereas in HF (blue dotted arrow) it is shifted to higher 
energies. Thus, aligning the chemical potentials, the position of the
BLA (TDGA) antibound state is in  disagreement (agreement) with the
exact result.~\cite{sei08} In addition, the TDGA gives a good account of
the low-energy spectral weight (see inset),
which is much larger and shifted to higher energies in the BLA. 
This was understood as due to the different way the self-energy is renormalized
by interactions in GA and in HF.~\cite{sei08} This dramatic difference
in performance gets greatly amplified for intersite correlations 
[see Fig.~\ref{figresu10}(b) and~(c)] because the band states acquire 
significant weight. 

The intersite correlations have also significant weight at the energy of
the s-wave antibound state $\omega'\sim U$. This can be understood from the 
second term of Eq.~\eqref{eq:pint} and  corresponds to processes in which an
intersite pair decays into an on-site pair. Notice that the d-wave pair cannot 
mix with a local s-wave pair at ${\bf q=0}$ but couples at finite  ${\bf q}$.

In addition, the exact result shows a satellite at $\omega'\approx 2U$
for the intersite cases corresponding to a process in which the two
particles are created at neighboring sites, that are initially single
occupied, thus leading to two on-site pairs in the final state. 
This can be visualized as the creation of the two particles in the
upper Hubbard band. Since this band is not present in our
starting point (i.e., GA), this satellite is absent in the pp-RPA. This feature 
reappears if one starts from a GA state that has both lower and upper Hubbard 
bands at the cost of spontaneous symmetry breaking. This case is analyzed 
in the following. 

\begin{figure}[htb]
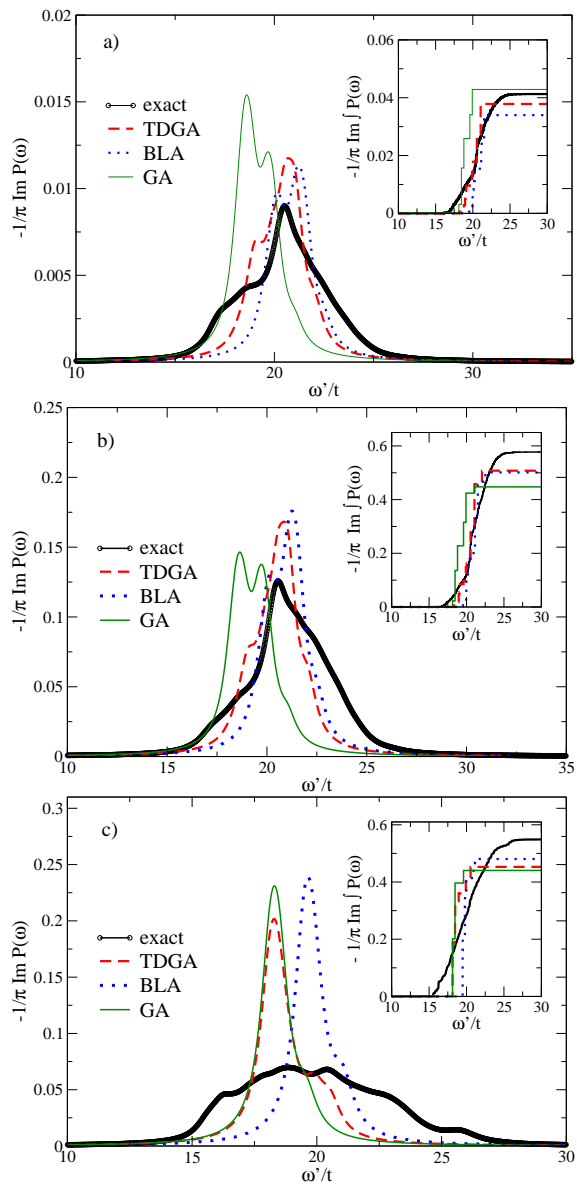

\includegraphics[width=7.5cm,clip=true]{fig6a.eps}
\includegraphics[width=7.5cm,clip=true]{fig6b.eps}
\includegraphics[width=7.5cm,clip=true]{fig6c.eps}
\caption{(color online) 
Imaginary part of the (two-particle addition) pairing correlation 
function Eq.~\eqref{eq:pcorr} for on-site pairing (a), extended s-wave (b) 
and d-wave (c) symmetry. Results are for a half-filled 18-site cluster and 
$U/t=10$. The underlying mean-field solution in BLA and TDGA is a 
spin-density wave. The delta-like excitations are convoluted by lorentzians 
with width $\epsilon = 0.5t$. The insets detail the frequency evolution of 
the two-particle spectral weight.}
\label{figresu10p18}
\end{figure}

We now consider the half-filled system where both HF and GA have a spin-density 
wave (SDW) ground state. The dynamical pairing correlations, for 18 
particles in the 18-site cluster, are shown in Fig.~\ref{figresu10p18}. 
Again, we are comparing absolute energies, but the position of $2\mu$ in the 
different approximations is aligned because the chemical potentials
coincide and are equal to the exact result $\mu=U/2$.

For local s-wave, extended s-save, and d-wave fluctuations the exact result has
a broad distribution of weight centered around $\omega'\sim 2U$. Clearly this 
corresponds to the high-energy satellite of the previous case. Now
practically all sites are singly occupied so the probability to find an
empty site where to create a local s-wave pair is very small and there
is practically no  weight at the energy of the antibound state
$\omega'\sim U$. Indeed, if the TDGA computation is done with a paramagnetic
state (which cannot reproduce the satellite at $2U$) the response
becomes identically zero.~\cite{sei08} This can be also seen from the
sum rule of Eqs.~\eqref{eq:intp1},~\eqref{eq:intp2} since in the paramagnetic 
TDGA and GA the double occupancy vanishes for the half-filled system above the
Brinkman-Rice transition point.

Breaking the symmetry and allowing for the SDW, both BLA and TDGA give a good 
estimate of the energy scale but with a more narrow distribution of weight. 
Notice that the on-site s-wave response has a much smaller
intensity consistent with the fact that the feature originates from
intersite excitations which get mixed with the on-site response. 

Processes in which the two particles are created on different singly occupied
sites are allowed in both approximate theories and they lead to
approximately similar results. In this case the interaction is
practically ineffective. In fact, the excitation spectra evaluated from $P^0$ 
(from the bare Gutzwiller Hamiltonian) shown with thin lines in 
Fig.~\ref{figresu10p18} differ only slightly from the TDGA result
and the latter only shows a small shift of spectral weight to higher
energies due to the inclusion of particle-particle correlations. 
This is in contrast to the paramagnetic case away form half filling 
(see Fig.~\ref{figresu10}) where all high-energy (antibound) states are
determined from poles in $P(\omega')$ but are absent in the non-interacting 
case $P^{0}(\omega')$ as discussed before.

The insets of Fig.~\ref{figresu10p18} show the evolution of the 
two-particle spectral weight. For both extended s-wave and d-wave
symmetry the total weight approaches $1/2$ in the pp-RPA theories  
in the limit $U\to \infty$ as for the AF solution of the two-site model 
in the right panel of Fig.~\ref{fig2s1}. This can be easily understood 
from the Neel limit. In contrast, the exact result has a larger weight
which can be understood from quantum fluctuations that induce spin flips.
In the extreme case of the two site model, the fluctuations lead to a local 
singlet and the exact spectral weight for $U\to \infty$ 
is twice the approximate one as discussed in connection with
Fig.~\ref{fig2s1}. Clearly this result is closely related to the
reduction of the magnetization in a quantum antiferromagnet. 

The structure of the TDGA interaction kernel Eq.~\eqref{espin} is more complex 
than the paramagnetic case. Fig.~\ref{figueff} displays
the corresponding non-vanishing elements as a function of $U/t$. 
For small on-site repulsion the local contribution behaves as
$V_{ii,ii}\approx U$, independent of the filling and the ground state,
as it should. With the onset of the SDW at $U/t \approx 4.1$, $V_{ii,ii}$
(dashed line) starts to deviate from the corresponding
interaction in the paramagnetic limit (full line) which diverges
at the Brinkmann-Rice transition.~\cite{sei08}
Interestingly, the interaction between local pairs on adjacent sites
(upper left inset of Fig.~\ref{figueff}) is always attractive 
in the SDW phase with a maximum attraction at $U/t \approx 10$.
In addition one finds an attractive or repulsive interaction between
a local pair on site $i$ and an intersite pair 
$c_{i,\uparrow} c_{j,\downarrow}$. For $i$ and $j$ nearest neighbors the
interaction is attractive (repulsive) if the pair has the same (opposite) 
spin with respect to the underlying Neel magnetic moments of sites $i$ and $j$. 
Nevertheless, these additional fluctuations in the TDGA cannot overcome
the strong residual on-site repulsion so that the system remains stable
against a transition towards superconductivity. This also holds
for a SDW ground state away from half filling.

\begin{figure}[htb]
\includegraphics[width=7cm,clip=true]{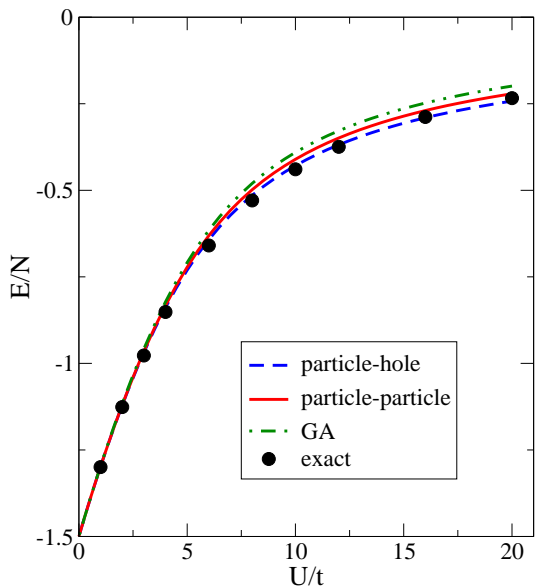}
\caption{(color online) 
Ground state energy of the half-filled $4 \times 4$ Hubbard model computed 
within GA, TDGA in the particle-hole channel, TDGA in the particle-particle 
channel, and exact diagonalization.}
\label{fig2s4}
\end{figure}

Finally, as in the two-site example, we calculate the energy correction
from the TDGA for a half filled $4\times 4$ cluster
from Eqs.~\eqref{eq:intp1} and~\eqref{eq:coupl}. In Fig.~\ref{fig2s4}, we 
compare the corresponding result for the particle-particle and particle-hole 
channel (from Ref.~\onlinecite{GOETZ1}) with the bare GA and the exact 
ground-state energy. The underlying saddle-point of the GA solution is a SDW.
As can be seen from Fig.~\ref{fig2s4}, the particle-hole TDGA correction is
approximately twice that in the particle-particle channel. The former
gives a quite accurate approximation for intermediate values of the
on-site repulsion but tends to overshot the exact result for large $U/t$
(note that these energy corrections are not derived from a 
variational principle
and thus do not constitute an upper bound for the exact result).
With the considered range of $U/t$ the particle-particle corrections always
are slightly higher in energy than the exact ground state.
However, in comparison with the HF+RPA energy corrections which are by far
too large,~\cite{GOETZ1} the TDGA yields a reasonable approximation to $E_0$
in both particle-hole and particle-particle channel.

\section{Conclusions}\label{sec:conc}

In this paper, we have extended the TDGA towards the inclusion
of pair correlations in the Hubbard model.
The present analysis is complementary to previous computations in the
particle-hole channel~\cite{GOETZ1,GOETZ2,goetz03} where we have analyzed
the spectrum of charge- and spin excitations.
In comparison with exact diagonalization results the TDGA yields a very good 
agreement for the dynamical pair correlation function, especially for the 
low-energy excitations and away from half filling where it performs
significantly better than the HF based BLA theory.

Compared to numerical methods~\cite{LANCZOS} our approach can be pushed to 
much larger systems. In particular, it is suitable for the evaluation of pair 
correlations in the negative-$U$ Hubbard model, where the approach is at least
qualitatively capable to capture the crossover from weak coupling BCS to
strong coupling behavior.~\cite{falk}
An important outcome of this investigation concerns the justification
of the antiadiabatic assumption for the time-evolution of the double
occupancy, which was needed in the charge particle-hole channel.

For the Hubbard Hamiltonian with particle-hole symmetry (e.g., nearest-neighbor
hopping at half filling), the superconducting instability and the 
charge-density wave instability in the particle-hole channel are
degenerate. Evaluation of the latter instability within the TDGA requires
to invoke the antiadiabaticity condition whereas the expansion in the
particle-particle channel of Eq.~\eqref{espin} goes without it.
The fact that the two calculations within the TDGA for the particle-hole 
and the particle-particle sector give the correct degeneracy of the 
instabilities for a charge-rotational invariant system clearly indicates
that the antiadiabaticity assumption was indeed correct, i.e., other
possibilities, as keeping the double occupancy fixed at the
stationary value (rather than to follow the time evolution of the
density matrix), would have led to an unphysical breaking of
charge-rotational symmetry.

Our approach does not produce a superconducting instability in the
Hubbard model. This is because, in the paramagnetic phase, the effective
interaction does not contain the attractive part due to the exchange
of spin fluctuations. Indeed, the superexchange scale $J$ is absent in
the paramagnetic GA. On the other hand, it turned out that the 
effective interaction in the SDW phase contained attractive contributions
between nearest-neighbor pairs which, however, are too weak in
order to yield a superconducting instability once the SDW is present.

The flexibility of the present approach allows one to study collective modes 
in inhomogeneous superconducting states once the Hamiltonian is augmented 
with a suitable pairing potential. This may find application in the physics 
of high-$T_c$ cuprates, where in many compounds the occurrence of electronic 
inhomogeneities (e.g., in the form of stripes) is now well established. 
Another possible application is in the field of ultra-cold atoms where the 
ground state is intrinsically inhomogeneous due to the presence of the 
confining parabolic potential. In addition, recent studies incorporate 
disorder to produce a glassy state, which may lead to similar physics as in 
the cuprates.~\cite{fall06}

\acknowledgments
G.S. acknowledges financial support from the Deutsche Forschungsgemeinschaft.
F.B. and J.L. partial support by CNR-INFM.

\appendix 

\section{Derivation of the charge-rotationally invariant GA
within the slave-boson approach}\label{AP}

The procedure implemented in the following  essentially consists of three
steps. Assume that in our initial reference frame we have non-vanishing
superconducting order, i.e., $\langle J_i^+\rangle \ne 0$ and (or) in
$\langle J_i^-\rangle \ne 0$. First, we rotate them locally to a new frame
where these expectation values vanish, i.e., $\langle \tilde J_i^+\rangle =
\langle \tilde J_i^-\rangle= 0$. This allows, as a second step, the
introduction of slave-bosons and associated fermions $\tilde f_{i\sigma}$ 
within the Kotliar-Ruckenstein scheme. For the bosons, we apply the
saddle-point (mean-field) approximation. Finally, in a third step, we rotate 
the fermions back to the original reference frame.

We define the local rotations in charge space by the following transformations
\begin{equation}\label{eq:tr1}
\widetilde{\bf \Psi}_i = {\bf U}_i^\dagger {\bf \Psi}_i \,\,\,\,\,\,\,
\widetilde{\bf \Psi}_i^\dagger =  {\bf \Psi}_i^\dagger {\bf U}_i,
\end{equation}
where
\begin{equation}
{\bf U}_i = \cos(\varphi_i/2) {\bf 1} + \imath 
\sin(\varphi_i/2)\vec{\boldsymbol{\tau}} \cdot \vec{\boldsymbol{\eta}},
\end{equation}
and $\vec{\boldsymbol{\eta}}= (\eta_x,\eta_y,0)$ is the rotation axis of 
length unity. The inverse transformation reads as
\begin{equation}\label{eq:inv}
{\bf \Psi}_i = {\bf U}_i \widetilde{\bf \Psi}_i \,\,\,\,\,\,\,
{\bf \Psi}_i^\dagger =  \widetilde{\bf \Psi}_i^\dagger {\bf U}_i^\dagger.
\end{equation}

Within the first step of our procedure we have the requirement
that the transformed pseudospin vector is given by
$\vec{\tilde J}_i=(0, 0, \tilde J_i^z)$. Applying the transformation of
Eq.~\eqref{eq:inv} to this vector one obtains the following relations
\begin{eqnarray}
J_i^x &=& - \eta_y \sin(\varphi_i) \tilde J_i^z, \\
J_i^y &=& \eta_x \sin(\varphi_i) \tilde J_i^z, \\
J_i^z &=& \cos(\varphi_i) \tilde J_i^z . \label{eq:jz}
\end{eqnarray}

Note that the spin operator
\begin{equation}
S_i^z  = \frac{1}{2} \left(  {\bf \Psi}_i^\dagger {\bf \Psi}_i  -1 \right) = 
\tilde S_i^z 
\end{equation}
is unchanged by the transformation.

Since by definition off-diagonal order vanishes in the rotated
frame we can now, as a second step, apply the Kotliar-Ruckenstein
slave-boson scheme:
\begin{equation}
\tilde c_{i\sigma} = z_{i\sigma}\tilde f_{i\sigma} \,\,\,\,\,\,
\tilde c_{i\sigma}^\dagger = z_{i\sigma}^\dagger\tilde f_{i\sigma}^\dagger,
\end{equation}
with
\begin{eqnarray}\label{zfak}
&& z_{i\sigma}=\\
&&\frac{1}{\sqrt{e_i^\dagger e_i + p_{i,-\sigma}^\dagger 
p_{i,-\sigma}}}
\left[ e_i^\dagger p_{i\sigma} + p_{i,-\sigma}^\dagger d_i \right]
\frac{1}{\sqrt{d_i^\dagger d_i + p_{i,\sigma}^\dagger p_{i,\sigma}}}.\nonumber
\end{eqnarray}
The double ($d$), singly ($p_\sigma$), and empty ($e_i$) occupancy
bosons are constrained by the following relations:
\begin{eqnarray}
\sum_\sigma p_{i,\sigma}^\dagger p_{i,\sigma}+ 2 d_i^\dagger d_i
&=& 2 \tilde J_i^z +1, \label{eq:c1} \\
p_{i,\uparrow}^\dagger p_{i,\uparrow} - p_{i,\downarrow}^\dagger p_{i,\downarrow}
&=& 2 \tilde S_i^z = 2 S_i^z, \label{eq:c2} \\
d_i^\dagger d_i + \sum_\sigma p_{i,\sigma}^\dagger p_{i,\sigma}
+ e_i^\dagger e_i &=& 1. \label{eq:c3}
\end{eqnarray}

Since we follow essentially a Gutzwiller-type approach, we now apply the 
mean-field approximation for the bosons. With the help of 
Eqs.~\eqref{eq:jz},~\eqref{eq:c1},~\eqref{eq:c2} and~\eqref{eq:c3}
we can eliminate all bosons but $d_i$ and express them via expectation
values in the {\it original} reference frame and $d_i^2$. One finds
\begin{eqnarray}
e_i^2 &=& -\frac{2}{\cos(\varphi_i)} \langle J_i^z \rangle +d_i^2, \\
p_{i\sigma}^2 &=& \frac{1}{2} + \frac{1}{\cos(\varphi_i)}\langle J_i^z \rangle
+ \sigma \langle S_i^z \rangle -d_i^2,
\end{eqnarray}
and $\sigma \equiv \pm 1$ in the latter equation.

Summarizing the steps we have performed so far the original 
fermion operators $c_{i\sigma}$ are related to the
Kotliar-Ruckenstein transformed ones $\tilde f_{i\sigma}$ in the
rotated frame via the transformation
\begin{equation}\label{eq:s12}
{\bf \Psi}_i = \left(\begin{array}{c} c_{i\uparrow} \\
c_{i\downarrow}^\dagger \end{array}\right)
= {\bf W}
\left(\begin{array}{c} \tilde f_{i\uparrow} \\
\tilde f_{i\downarrow}^\dagger \end{array}\right),
\end{equation}
with
\begin{equation}
{\bf W}=\left(\begin{array}{cc}
z_{i,\uparrow}\cos\frac{\varphi_i}{2} & \imath(\eta_x - i\eta_y) 
\sin\frac{\varphi_i}{2} z_{i,\downarrow} \\
\imath(\eta_x + i\eta_y) \sin\frac{\varphi_i}{2} z_{i,\uparrow} & 
z_{i,\downarrow}\cos\frac{\varphi_i}{2} \end{array}\right).
\end{equation}

Finally we transform the fermion operators $\tilde f_{i\sigma}$
back to the original frame [see Eq.~\eqref{eq:tr1}]
\begin{equation}
\left(\begin{array}{c} \tilde f_{i\uparrow} \\
\tilde f_{i\downarrow}^\dagger \end{array}\right)
= {\bf U}
\left(\begin{array}{c} f_{i\uparrow} \\
 f_{i\downarrow}^\dagger \end{array}\right),
\end{equation}
so that the charge-rotational invariant Gutzwiller representation of 
the fermions is given by
\begin{equation}
{\bf \Phi}_i = \left(\begin{array}{c} f_{i\uparrow} \\
f_{i\downarrow}^\dagger \end{array}\right)
= {\bf W U}
\left(\begin{array}{c} \tilde f_{i\uparrow} \\
\tilde f_{i\downarrow}^\dagger \end{array}\right).
\end{equation}
The complete transformation matrix ${\bf A}={\bf W U}$ reads as:
\begin{equation}\label{amatap}
{\bf A}_i= \left( \begin{array}{cc} 
z_{i\uparrow}\cos^2\frac{\varphi_i}{2}+z_{i\downarrow}
\sin^2\frac{\varphi_i}{2} & 
\frac{\langle J_i^-\rangle }{2 \langle J_i^z \rangle } [z_{i\uparrow}-z_{i\downarrow}]\cos\varphi_i \\ 
\frac{\langle J_i^+  \rangle}{2\langle J_i^z \rangle }[z_{i\uparrow}-z_{i\downarrow}]\cos\varphi_i& 
z_{i\uparrow}\sin^2\frac{\varphi_i}{2}+z_{i\downarrow}\cos^2\frac{\varphi_i}{2}
\end{array} \right),
\end{equation}
with
\begin{equation}
\tan^2\varphi_i=\frac{\langle J_i^+\rangle \langle J_i^-\rangle }{ \langle J_i^z \rangle^2},
\end{equation}
and $J_i^+$, $J_i^-$ and $J_i^z$ are defined as in
Eqs.~\eqref{eqjz},~\eqref{eqjp}, and~\eqref{eqjm} but with the pseudofermion 
operators $f_{i\sigma}$ instead of $c_{i\sigma}$. 

Note the formal similarity of Eq.~\eqref{amat} with the corresponding
transformation of the spin-rotation invariant Gutzwiller 
approximation.~\cite{goetz03}

We now turn to the transformation of the interaction term of the
Hubbard Hamiltonian which has to be performed within the second
step of the above scheme but {\it before} the saddle-point approximation
has been applied to the bosons.
Rewriting the interaction in terms of the transformation Eq.~\eqref{eq:s12}
and obeying the constraints Eqs.~\eqref{eq:c1}-\eqref{eq:c3} we obtain
\begin{equation}\label{eq:d}
n_{i\uparrow} n_{i\downarrow} = \cos^2\frac{\varphi_i}{2} d_i^\dagger d_i
+ \sin^2\frac{\varphi_i}{2} e_i^\dagger e_i,
\end{equation}
which now can be again treated in the mean-field approximation. 

We finally obtain for the charge-rotational invariant Gutzwiller
energy functional of the Hubbard model

\begin{eqnarray}
E&=& \sum_{i,j}
t_{ij} \langle{\bf \Phi}_i^\dagger {\bf A}_{i} \boldsymbol{\tau}_z
{\bf A}_{j}{\bf \Phi}_j\rangle \nonumber \\
&+& U\sum_{i}\left[d_i^2 - J_i^z (\sqrt{1+\tan^2\varphi_i}-1)\right]. \label{EGAP}
\end{eqnarray}
The term multiplying $U$ is clearly the sum of the  double occupancies
$D_i$ of the original hamiltonian. Therefore, we perform the substitution
\begin{equation}
D_i = d_i^2- J_i^z (\sqrt{1+\tan^2\varphi_i}-1)
\end{equation}
to reach Eq.~\eqref{EGA}. 
With this definition it is also straightforward to proof the equivalence
of the saddle-point z-factors Eq.~\eqref{zfak} with the Gutzwiller
renormalization factors $z_{i\sigma}$ from Eq.~\eqref{qfak} and of
the two representations of the transformation matrix ${\bf A}$ in
Eq.~\eqref{amat} and Eq.~\eqref{amatap}.

\section{Derivatives of the hopping factor}\label{APA}

The derivatives appearing in Eqs.~\eqref{z1} and~\eqref{z2} are given by:
\begin{widetext}
\begin{eqnarray}
A_i' &=& \left. \frac{\partial A_{i+-}}{\partial \langle J_i^-\rangle}\right|_0 =
\left. \frac{\partial A_{i-+}}{\partial  \langle J_i^+\rangle} \right|_0 
= \frac{z_{i\uparrow} - z_{i\downarrow}}{2 J_i^z} \\
A_{i\tau\tau}^{''} &=& \left. \frac{\partial^2 A_{i\tau\tau}}
{\partial \langle J_i^+\rangle \partial  \langle J_i^-\rangle}\right|_0   =
-\sigma \frac{z_{i\uparrow}-z_{i\downarrow}}{(1-n_i)^2} +
\frac{z_{i\tau}}{n_i-1}\frac{n_{i\tau}-\frac{1}{2}}{n_{i\tau}(1-n_{i\tau})}
\\
&+&\frac{1}{1-n_i}\frac{1}{\sqrt{n_{i\tau}(1-n_{i\tau})}}
\left\lbrace \sqrt{\frac{n_{i\tau}-D_i}{1-n_i+D_i}}-\frac{1}{2}
\sqrt{\frac{1-n_i+D_i}{n_{i\tau}-D_i}}-\frac{1}{2}\sqrt{\frac{D_i}{n_{i,-\tau}-D_i}}\right\rbrace.
\end{eqnarray}
Notice that the index $\tau$ for the matrix element of 
${\bf A}$ appears at the place of the spin index on the right. In this
case one should interpret  $+=\uparrow $ and $-=\downarrow $.

In case of a homogeneous, paramagnetic saddle point these
expressions simplify to
\begin{eqnarray}
A_{i}' &=& 0 \\
A_{i\tau\tau}^{''}&=& \frac{2 z_0}{n(2-n)} + \frac{z_0}{1-n}
\left\lbrack \frac{1}{1-n+D + \sqrt{D(1-n+D)}} - \frac{1}{n-2D}\right\rbrack
\end{eqnarray}
\end{widetext}
where $z_0$ denotes the spatially homogeneous Gutzwiller factor defined 
in Eq.~\eqref{qfac}.

\section{Sum rule for intersite pairing operators}\label{APB}

We define local and extended s-wave pairing operators on 
a hypercubic lattice (coordination number ${\cal Z}$) as
\begin{eqnarray*}
\Delta_i^s &=& c_{i\downarrow}c_{i\uparrow} \\
\Delta_i^{es} &=& \sum_{j}\gamma_{ij} 
\left[ c_{i\downarrow}c_{j\uparrow} + c_{j\downarrow}c_{i\uparrow} \right]
\end{eqnarray*}
where $\gamma_{ij}=1$ for $ij$ nearest neighbors and  
$\gamma_{ij}=0$ otherwise.
We consider the Hubbard model Eq.~\eqref{HM} with nearest-neighbor hopping 
only. The kinetic energy operator can then
be represented as
\begin{equation}
T=-t\sum_{ij\sigma}\gamma_{ij} c^\dagger_{i\sigma}c_{j\sigma}
\end{equation}
and the commutator of $\Delta_i^l$ with $H$ yields
\begin{equation}
\left[ H, \Delta_i^l\right] = t \Delta_i^{es} - U \Delta_i^l .
\end{equation}
For the double-commutator with $(\Delta_i^{es})^{\dagger}$ we obtain
\begin{eqnarray} 
&&\left[(\Delta_i^{es})^{\dagger}, \left[H, \Delta_i^l\right]\right]
= -2{\cal Z} t + {\cal Z} t (n_{i\uparrow} + n_{i\downarrow}) \nonumber \\
&+& t \sum_{nm}\gamma_{in}\gamma_{im}\left[c_{n\uparrow}^\dagger
  c_{m\uparrow} + c_{n\downarrow}^\dagger c_{m\downarrow} \right]\nonumber \\
&-& U \sum_{n}\gamma_{in}\left[c_{n\uparrow}^\dagger c_{i\uparrow} +
c_{n\downarrow}^\dagger c_{i\downarrow} \right] \label{doubcomm}
\end{eqnarray}
which only depends on one-particle operators.

Taking the expectation value of Eq.~\eqref{doubcomm} and inserting a complete
set of $N+2$ and $N-2$ particle states we find the  following sum rule
\begin{eqnarray}
&& \sum_i \langle \left[(\Delta_i^{es})^{\dagger}, \left[H,
    \Delta_i^l\right]\right]\rangle \nonumber\\
&=& \sum_{i,\nu}\omega^+_\nu \langle \Psi_0^N|\Delta_i^l|\Psi_\nu^{N+2}\rangle
\langle\Psi_\nu^{N+2}|(\Delta_i^{es})^{\dagger}|\Psi_0^N\rangle \nonumber \\
&-& \sum_{i,\nu}\omega^-_\nu \langle
    \Psi_0^N|(\Delta_i^{es})^{\dagger}|\Psi_\nu^{N-2}\rangle
\langle\Psi_\nu^{N-2}|\Delta_i^l|\Psi_0^N\rangle \nonumber \\
&=& -2 N_L {\cal Z} t\left( 1   - \langle n\rangle \right) +\frac{U}{t}\langle T
    \rangle \nonumber \\
&+& t \sum_{in\ne m}\gamma_{in}\gamma_{im}\left\langle\left[c_{n\uparrow}^\dagger
  c_{m\uparrow} + c_{n\downarrow}^\dagger c_{m\downarrow} \right]\right\rangle \label{eq:sr}
\end{eqnarray}
where $N_L$ denotes the number of  lattice
sites, $\langle n\rangle$ the particle density and 
 $\omega^{\pm}_\nu=\pm [E_\nu(N\pm2)-E_0(N)]$.

In case of the time-dependent GA the states $|\Psi_\nu^{N\pm2}\rangle$ 
are interpreted as Gutzwiller projected RPA states, i.e. they
incorporate Gutzwiller and particle-particle correlations. In the
spirit of the GA matrix elements are evaluated in term of the unprojected RPA
states as  
$$\langle\Psi_0^N|(\Delta_i^{es})^{\dagger}|\Psi_\nu^{N-2}\rangle=
\langle0,N|(\bar \Delta_i^{es})^{\dagger}|\nu,{N-2}\rangle$$ where 
$\bar \Delta_i^{es}$ is renormalized with the z factors as in
Eq.~\eqref{exswave}. As in HF+RPA theories one expect that the r.h.s
is equal to the l.h.s when evaluated at the static mean field level in
this case the GA. Indeed evaluating the kinetic expectation values on
the GA approximation (incorporating the z factors) one finds and
identity as expected. 

We can explicitely demonstrate the procedure for the half-filled
two-site model where local and intersite pairing operators are defined in 
Eqs.~\eqref{local2s} and~\eqref{inter2s}. For this special case the sum rule
Eq.~\eqref{eq:sr} becomes
\begin{eqnarray}
\langle \left[(\Delta^{es})^{\dagger}, \left[H,
    \Delta^l\right]\right]\rangle
&=& \omega^+ \langle \Psi_0^2|\Delta^l|\Psi^{2+2}\rangle
\langle\Psi^{2+2}|(\Delta^{es})^{\dagger}|\Psi_0^2\rangle \nonumber \\
&-& \omega^- \langle\Psi_0^2|(\Delta^{es})^{\dagger}|\Psi^{2-2}\rangle
\langle \Psi^{2-2}|\Delta^l|\Psi_0^2\rangle \nonumber \\
&=& \frac{U}{2t}\langle T \rangle .
\label{eq:sr2s}
\end{eqnarray}
 For the l.h.s. of Eq.~\eqref{eq:sr2s} we find from Eqs.~\eqref{eq:matl},
and~\eqref{eq:matis}
\begin{eqnarray}
&&\omega^+ \langle 0,2|\Delta^l|2+2\rangle
\langle 2+2|(\Delta^{es})^{\dagger}|0,2\rangle \nonumber \\
&-& \omega^- \langle 0,2|(\Delta^{es})^{\dagger}|2-2\rangle
\langle 2-2|\Delta^l|0,2\rangle \nonumber \\
&=& -\frac{1}{2}\omega^+\left[(X^+)^2-(Y^+)^2\right]
+\frac{1}{2}\omega^-\left[(X^-)^2-(Y^-)^2\right]\nonumber \\
&=& -\frac{1}{2} \left[\omega^+ + \omega^-\right]=-2\Sigma_0=-U .
\end{eqnarray}
Evaluating the r.h.s. of Eq.~\eqref{eq:sr2s} within the GA one obtains
\begin{equation}
\frac{U}{2t}\langle T \rangle_{GA} = \frac{U}{2t} (-2t)(1-u^2) = -U(1-u^2)
\end{equation}
so that in the time-dependent GA the intersite pairing operator  
Eq.~\eqref{inter2s} has to be renormalized according to 
\begin{equation}
\bar\Delta_x = \frac{1}{\sqrt{2}}(1-u^2)(c_{1\downarrow}c_{2\uparrow} 
- c_{1\uparrow}c_{2\downarrow})
\end{equation}
in order to fulfill the sum rule Eq.~\eqref{eq:sr2s}.

\end{document}